\documentclass[english]{iopart}

\begin{document}

\title[Series expansion method and ODE of $\chi^{(4)}$ ]
{\Large
Ising model susceptibility: Fuchsian differential equation for $\chi^{(4)}$
 and its factorization properties}
 
\author{ 
N. Zenine$^\S$, S. Boukraa$^\dag$, S. Hassani$^\S$ and
J.-M. Maillard$^\ddag$}
\address{\S  Centre de Recherche Nucl\'eaire d'Alger, \\
2 Bd. Frantz Fanon, BP 399, 16000 Alger, Algeria}
\address{\dag Universit\'e de Blida, Institut d'A{\'e}ronautique,
 Blida, Algeria}
\address{\ddag\ LPTL, Universit\'e de Paris 6, Tour 24,
 4\`eme \'etage, case 121, \\
 4 Place Jussieu, 75252 Paris Cedex 05, France} 
\ead{maillard@lptl.jussieu.fr,  sboukraa@wissal.dz, njzenine@yahoo.com}

\begin{abstract}
We give the Fuchsian linear differential equation satisfied by $\chi^{(4)}$,
the ``four-particle'' contribution to the
 susceptibility of the isotropic square lattice Ising model.
This Fuchsian differential equation is deduced from a series expansion method 
introduced in two previous papers and is applied with some symmetries and
tricks specific to $\chi^{(4)}$.
The corresponding order ten linear differential
 operator exhibits a large set of factorization properties. Among 
these factorizations one is highly remarkable: it corresponds to the fact that
the two-particle contribution $\chi^{(2)}$ is actually a solution of this 
order ten linear differential  operator. This result, together with 
a similar one for the order seven differential  operator
corresponding to the three-particle contribution, $\chi^{(3)}$,
leads us to a conjecture on the structure of all the $\, n$-particle
contributions $\,\chi^{(n)}$. 
\end{abstract} 
\vskip .5cm


\noindent {\bf AMS Classification scheme numbers}: 34M55, 47E05, 81Qxx, 32G34, 34Lxx, 34Mxx, 14Kxx


\section{Introduction}

The magnetic susceptibility of square lattice Ising model,
can be written \cite{wu-mc-tr-ba-76} as an infinite sum
\begin{eqnarray}
\chi (T)\,=\,\,\sum_{n=1}^{\infty }\chi ^{(n)}(T) 
\end{eqnarray}
of individual contributions,
with the odd (respectively even) $n$ corresponding to 
high (respectively low) temperature case.
These individual contributions are 
$(n-1)$-dimensional integrals~\cite
{nappi-78,pal-tra-81,yamada-84,yamada-85,nickel-99,nickel-00}, and are
seen as successive  $n-$particle contributions to the susceptibility \cite{wu-mc-tr-ba-76}.

To get an understanding of the analytical structure of $\chi$, two
approaches are usually taken nowadays. One approch taking into account a
 fundamental non-linear symmetry,
namely non-linear Painlev\'{e} difference
 equations~\cite{or-ni-gu-pe-01b,or-ni-gu-pe-01,coy-wu-80,perk-80,jim-miw-80,coy-wu-81}.
provides a series expansion for the whole susceptibility $\,\chi $.
With this method coefficients of $\chi$ were recently generated
\cite{or-ni-gu-pe-01b}.

The second approach considers the individual $n$-particle excitations as given
by $(n-1)$-dimensional integrals. Isotropic series coefficients are generated
\cite{nickel-99,nickel-00,or-ni-gu-pe-01b}. This latter method 
 allows one to seek the differential equations satisfied  by the $\chi^{(n)}$,
 since they are $\, D$-finite contrary to the whole susceptibility $\chi$
for which there are strong indications that it has a natural boundary in the
complex plane of the variable
$\,s\,=sh(2K)$, where $K=J/kT$ is the usual Ising model coupling
constant.
This has been shown for the isotropic case \cite{nickel-99} and
for the anisotropic case \cite{HaMa88,ha-ma-oi-ve-87,gut-ent-96}.
Such a function cannot be $\, D$-finite.

The understanding of the magnetic susceptibility may then require
the knowledge of each (or some) of the individual contributions. This knowledge can be in the
form of a closed expression, as is the case for $\chi^{(1)}$ and $\chi ^{(2)}$ or in the
form of a differential equation as found for $\chi^{(3)}$ \cite{ze-bo-ha-ma-04,ze-bo-ha-ma-05}.
The last case was far from being obvious and has required the building of an
original method of expansion.
The use of a remarkable formula allowed us to give the series expansion in the
temperature variable (or a closely related variable)
where the $(n-1)$-dimensional integrals have been fully performed.

In this paper, we continue to use this expansion method to tackle the next
individual contribution, namely, $\chi^{(4)}$. We should note that, although,
the method is general and applicable for high or low temperature and for
any $n$, some of the tricks and tools used may be specific
for a given  $\chi^{(n)}$.
 In Section 2, we present the basic
features of the expansion method that allow us to obtain the fully
integrated $\,\chi ^{(4)}$ as four sums of products of four hypergeometric
functions, without any numerical approximation. In Section 3, we give the
homogeneous Fuchsian linear differential equation satisfied by $\,\chi ^{(4)}$.
Section 4 contains some remarkable algebraic properties of this differential
equation. Finally, Section 5 contains our conclusions.

\section{Fully integrated $\tilde{\chi}^{(4)}$ expansion}
\subsection{The expansion method}

Let us focus on the fourth contribution to the susceptibility $\chi $
defined by the triple integral as given in \cite{nickel-00} 
\begin{eqnarray}
\label{def KHI4Tilda}
\chi^{(4)} &=&(1-s^{-4})^{1/4}\,\cdot \,\tilde{\chi}^{(4)}  \nonumber \\
\tilde{\chi}^{(4)} &=&\int_{0}^{2\pi }\frac{d\phi _{1}}{2\pi }\int_{0}^{2\pi
}\frac{d\phi _{2}}{2\pi }\,\int_{0}^{2\pi }\frac{d\phi _{3}}{2\pi }\cdot
\tilde{y}_{1}\tilde{y}_{2}\tilde{y}_{3}\tilde{y}_{4}\cdot R^{(4)}\cdot H^{(4)}  
\end{eqnarray}
with
\begin{equation}
\label{defR4}
R^{(4)}\,=\,{\frac{1+\tilde{x}_{1}\tilde{x}_{2}\tilde{x}_{3}\tilde{x}_{4}}
{1-\tilde{x}_{1}\tilde{x}_{2}\tilde{x}_{3}\tilde{x}_{4}}}
\end{equation}

\begin{equation}
\label{defH4}
H^{(4)}\,=\frac{1}{4!}\,\prod_{i<j}\frac{\tilde{x}_{i}\tilde{x}_{j}}
{(1-\tilde{x}_{i}\tilde{x}_{j})^{2}}
\cdot (Z_{i}-Z_{j})^{2}  
\end{equation}

\begin{equation}
\label{defZ}
Z_{n}\,=\,\exp (i\phi _{n}), \quad \quad n=1, \cdots, 4  
\end{equation}
\begin{eqnarray}
\label{contrainte}
\phi _{1}+\phi _{2}+\phi _{3}+\phi _{4}=0  
\end{eqnarray}
\begin{eqnarray}
\label{varXY}
\tilde{x}_{n} &=&{\frac{s}{1+s^{2}-s\cos {\phi }_{{n}}+\sqrt{(1+s^{2}-s\cos {\phi
_{n}})^{2}-s^{2}}}},   \\
\tilde{y}_{n} &=&{\frac{s}{\sqrt{(1+s^{2}-s\cos {\phi _{n}})^{2}-s^{2}}}},\quad \quad
n=1, \cdots, 4 \quad  \nonumber
\end{eqnarray}

Instead of the variable $s$, we found  it more suitable to use $w={\frac{{1}}{%
{2}}}s/(1+s^{2})$ which has, by construction, Kramers-Wannier duality
invariance ($s\,\leftrightarrow \,1/s$) and \ thus allows us to deal with both
limits (high and low temperature, small and large $s$)
 on an equal footing \cite{ze-bo-ha-ma-04,ze-bo-ha-ma-05}.
It is also convenient to consider the scaled variables
\begin{eqnarray}
x_{n} &=&\frac{\tilde{x}_{n}}{w}={\frac{2}{1-2w\cos {\phi _{n}}+\sqrt{(1-2w\cos {%
\phi _{n}})^{2}-4w^{2}}}},  \label{varxy} \\
y_{n} &=&\frac{\tilde{y}_{n}}{2w}={\frac{1}{\sqrt{(1-2w\cos {\phi _{n}})^{2}-4w^{2}}}%
}  \nonumber
\end{eqnarray}
which behave like $\, 1+O(w)$ at small $w$.

As performing the integrals in (\ref{def KHI4Tilda}) is highly non trivial,
we apply the ``expansion method'' previously
described in \cite{ze-bo-ha-ma-04,ze-bo-ha-ma-05}, where the key ingredient
was the Fourier expansion of $y\,x^{n}$, a quantity appearing in any
$\chi ^{(p)}$.

This remarkable formula, for $y\,x^{n}$, that carries only
\emph{one summation index} reads
\begin{eqnarray}
\label{Four-yxn}
yx^{n}\, &=& \,a(0,n)\,+2\sum_{k=1}^{\infty}w^{k}\,a(k,n)\,\cos {k\phi }  \nonumber \\
& = & \sum_{k=-\infty }^{\infty }\,A(k,n)\,Z^{k}=
\sum_{k=-\infty}^{\infty }A(k,n)\,Z^{-k}
\end{eqnarray}
with
\begin{equation}
\label{def-A}
A(k,n)\,=\,\,A(-k,n)\,=\,\,w^{\left\vert k\right\vert }\,
a(\left\vert k\right\vert ,n)  
\end{equation}
where $a(k,n)$ is a \emph{non terminating} hypergeometric series that reads
\begin{eqnarray}
\label{defa}
&& a(k,n)={m \choose k} \cdot   \\
&& {_{4}}F_{3}\Bigl({\frac{{(1+m)}}{{2}}},{\frac{{(1+m)}}{{2}}},{\frac{{(2+m)}%
}{{2}}},{\frac{{(2+m)}}{{2}}};1+k,1+n,1+m;16w^{2}\Bigr)  \nonumber
\end{eqnarray}
where $m=k+n$. Note that $a(k,n)=a(n,k)$.

The integrand of $\,\tilde{\chi}^{(4)}$ is expanded in the various variables 
$x_{j}$, instead of the variable $w$. In this framework, with the help of
the Fourier expansion (\ref{Four-yxn}), the angular integration becomes
straightforward as was shown in \cite{ze-bo-ha-ma-05} for the
$\tilde{\chi}^{(3)}$ case (see below for the $\tilde{\chi}^{(4)}$ case).

\subsection{Calculation of $\tilde{\chi}^{(4)}$}
\label{calcuKi4}
With $H^{(4)}$ taken as in (\ref{defH4}), $\tilde{\chi}^{(4)}$ will be given
by {\it five} summations on products of four hypergeometric functions.
This is shown in Appendix A.
This route is feasible for any $\chi^{(n)}$ and does not use any symmetry or
tricks specific to the considered $\chi^{(n)}$.

In the following, we use alternatively a
simplified form of $H^{(4)}$ equivalent, for integration purposes, to
(\ref{defH4}) such that $\tilde{\chi}^{(4)}$ will be expressed by
only {\it four} summations (i.e., one summation less compared to the
expression (\ref{KHI4T-form2}) given in Appendix A).
Let us just sketch the salient steps of this calculation.
The details are left to Appendices B, C and D.

With the help of the key relation
\begin{equation}
\label{relZX}
(Z_{i}-Z_{j})\,\frac{\tilde{x}_{i}\tilde{x}_{j}}{1-\tilde{x}_{i}\tilde{x}_{j}}\,
=\,\,-(\tilde{x}_{i}-\tilde{x}_{j})\,\frac{Z_{i}Z_{j}}{1-Z_{i}Z_{j}}  
\end{equation}
the quantity $H^{(4)}$ becomes\footnote[1]{
We have used the constraint (\ref{contrainte}) in the form $\prod_{i}Z_{i}$ $%
=1$.}:
\begin{eqnarray}
\label{H4form1}
H^{(4)}\,=\frac{1}{4!}\, \prod_{i<j}\frac{\tilde{x}_{i}-\tilde{x}_{j}}
{1-\tilde{x}_{i}\tilde{x}_{j}}  \cdot
\prod_{i<j}\frac{Z_{i}-Z_{j}}{1-Z_{i}Z_{j}}.
\end{eqnarray}
Using the symmetry of the rest of the integrand in the angular variables,
the quantity $H^{(4)}$ can be written (see Appendix B) as\footnote[2]{
Throughout this paper the notation $\equiv $ stands for equality for
integration purposes.}
\begin{eqnarray}
\label{H4 simple}
H^{(4)} &\equiv &-\frac{1}{8}(Z_{1}-Z_{2})(Z_{3}-Z_{4})\frac{Z_{1}Z_{2}}{%
(1-Z_{1}Z_{2})^{2}}\cdot  \nonumber \\
&&\Bigl( P_{12-34}^{(x)}-\left( 1\leftrightarrow 3\right) -\left(
2\leftrightarrow 3\right) \Bigr)  
\end{eqnarray}
where :
\begin{eqnarray}
\label{PX12-34}
P_{12-34}^{(x)} &=&\, (\tilde{x}_{1}-\tilde{x}_{2})(\tilde{x}_{3}-\tilde{x}_{4})\cdot
 \\
&&\Bigl( \frac{\tilde{x}_{1}\tilde{x}_{2}}{1-\tilde{x}_{1}\tilde{x}_{2}}
\frac{\tilde{x}_{3}\tilde{x}_{4}}{1-\tilde{x}_{3}\tilde{x}_{4}}+
\frac{(\tilde{x}_{1}\tilde{x}_{2})^{2}}{1-\tilde{x}_{1}\tilde{x}_{2}}
+\frac{(\tilde{x}_{3}\tilde{x}_{4})^{2}}{1-\tilde{x}_{3}\tilde{x}_{4}}
\Bigr).  \nonumber
\end{eqnarray}
Taking the expression (\ref{H4 simple}) for $H^{(4)}$, expanding the
integrand in the $x_{j}$ variables, one notes that the integrand depends
only on combinations of the form
\begin{equation}
\label{formeintegrand}
\left( \prod_{i=1}^{4}y_{i}\cdot x_{i}^{n_{i}}\right) \cdot
(Z_{1}-Z_{2})(Z_{3}-Z_{4})\frac{Z_{1}Z_{2}}{(1-Z_{1}Z_{2})^{2}}
\end{equation}
which have simple integration rules (see Appendix C) and, thus, the problem
of angular integrations is settled.

Finally, the expansion method described above, together with the form (\ref
{H4 simple}) of $H^{(4)}$, allow us to obtain $\tilde{\chi}^{(4)}(w)$ as a 
\textit{fully integrated expansion} in the form (see Appendix D)
\begin{eqnarray}
\label{KHI4T-FINAL}
\tilde{\chi}^{(4)} &=&16\, w^{16}\cdot \sum_{m=0}^{\infty }\sum_{k=0}^{\infty
}\sum_{n=0}^{\infty }\sum_{j=0}^{\infty }w^{8m+4k+4n+2j}\cdot
(2m+1)(2m+2k+1)  \nonumber \\
&&(1+\theta (j-1))\cdot (1+\theta (k-1))\cdot  \nonumber \\
&& {{1}\over{2}} \Bigl( V(m,m+k,n,n+j)+V(m,m+k,n+j,n) \Bigr)   
\end{eqnarray}
where $\theta (x)$ is the step function defined as
\begin{equation}
\label{defTHETA}
\theta (x)=\left\{ 
\begin{array}{ll}
1, & \quad x\geq 0 \\ 
0, & \quad x<0
\end{array}
\right.  
\end{equation}
and
\begin{eqnarray}
\label{defHYP-KHI4}
&& V(m,k,n,j) =d(m,m+1;j+k+2)\cdot d(n+m+2,n+m+3;k)  \nonumber \\
&&\qquad +d(m,m+1;k)\cdot d(n+m+2,n+m+3;j+k+2)  \nonumber \\
&&\qquad +d(m,n+m+3;j+k+2)\cdot d(m+1,n+m+2;k)  \nonumber \\
&&\qquad +d(m,n+m+3;k)\cdot d(m+1,n+m+2;j+k+2)  \nonumber \\
&&\qquad -d(m,n+m+2;j+k+2)\cdot d(m+1,n+m+3;k)  \nonumber \\
&&\qquad -d(m,n+m+2;k)\cdot d(m+1,n+m+3;j+k+2)  
\end{eqnarray}
with :
\begin{equation}
\label{def-d}
d(n_{1},n_{2};k)=\, a(n_{1},k+1)\,a(n_{2},k)-a(n_{1},k)\,a(n_{2},k+1).  
\end{equation}

\subsection{Series generation}
\label{sergener}
Note that the summand in equation (\ref{KHI4T-FINAL}) depends only on
combinations of $C_{d}(n_{1},n_{2},k,j)$ which is the coefficient of $w^{2j}$
in the expansion of $d(n_{1},n_{2};k)$. It is given by
\begin{eqnarray}
\label{defCd}
C_{d}(n_{1},n_{2},k,j) &=& \sum_{i=0}^{j} \, \Bigl(
C_{a}(n_{1},k+1,i)\,C_{a}(n_{2},k,j-i) \nonumber \\
&& \quad \quad -C_{a}(n_{1},k,i)\, C_{a}(n_{2},k+1,j-i) \Bigr)
\end{eqnarray}
where $C_{a}(n,k,i)$ is the coefficient of $w^{2i}$ in the expansion of
$a(n,k)$ which reads :
\begin{equation}
\label{defCa}
C_{a}(n,k,i)=\, {n+k+2i \choose i} {n+k+2i \choose i+k}  
\end{equation}
From our integrated form of $\tilde{\chi}^{(4)}$, the generation of series
coefficients becomes straightforward. \emph{Recall that} (\ref{KHI4T-FINAL}) \emph{is
already integrated}, and, thus, the computing time to obtain the series
coefficients comes from the evaluation of the sums. For the forms used, this
time is of order $N^{7}$. Improvements can be made by optimal data storage
in order to avoid repeated summation evaluations. Actually, we have found it more
efficient to store the coefficients $C_{d}$.

We have been able to generate, from formal calculations,
a long series of coefficients from the
expression (\ref{KHI4T-FINAL}) up to order\footnote{More precisely we obtained 
this series up to order 368 with a 2 Giga-memory computer, up to order
390 with a 3 Giga-memory computer, and up to  order 432
 with a 8 Giga-memory computer of the stix laboratory at
 the Ecole Polytechnique (medicis platform for formal calculations).} 432 : 
\begin{eqnarray}
\label{series}
{\frac{\tilde{\chi}^{(4)}(w)}{16\,w^{16}}}\,\,=\,\, 1\,+64\,{w}^{2}
+2470\,{w}^{4}\,+74724\,w^{6}\,\,+\, ... \,\, +O(w^{434}) 
\end{eqnarray}
Note that our formal calculation program has been rewritten by
J. Dethridge\footnote{Private  communication.} into an optimized
C++ program that can give the series in few hours using very
little memory.

\section{The Fuchsian differential equation satisfied by $\tilde{\chi}^{(4)}$}
\label{fuchsdiff}
 It is clear from the expression (\ref{KHI4T-FINAL}) that $\tilde{\chi}^{(4)}$%
 is even\footnote[3]{This is also the case for any 
$\tilde{\chi}^{(2n)}$ (see for instance~\cite{nickel-00}).} in $w$.
We thus introduce, in the following, the variable $x=16 \,w^2$.
With our long series, and with a dedicated program, we
have succeeded in obtaining the differential equation for
$\,\tilde{\chi}^{(4)}$ that is given by (with $x=16 \,w^2$)
\begin{eqnarray}
\label{fuchs}
\sum_{n=1}^{10}\,a_{n}(x)\cdot {\frac{{d^{n}}}{{dx^{n}}}}F(x)\,\,=\,\,\,\,\,0
\end{eqnarray}
with  
\begin{eqnarray} 
\label{defQ}
&&a_{10}= -512\,{x}^{6} \left(x-4 \right)  \left(1-x \right)^{6} P_{10}(x),  \nonumber \\
&&a_{9}= 256\, \left(1-x \right)^{5}{x}^{5} P_{9}(x),
\quad a_{8}= -384\, \left(1-x \right)^{4}{x}^{4} P_{8}(x),  \nonumber \\
&&a_{7}= 192\, \left(1-x \right)^{3}{x}^{3} P_{7}(x),\qquad 
a_{6}= -96\, \left(1-x \right)^{2}{x}^{2} P_{6}(x),  \nonumber \\
&&a_{5}= 144\, \left(1-x \right)\, x \,P_{5}(x),\qquad a_{4}= -72 P_{4}(x),
\quad a_{3}= -108 P_{3}(x) \nonumber \\
&& a_2= -54 P_2(x) \qquad a_1= -27 P_1(x)
\end{eqnarray}
where $P_{10}(x),P_{9}(x)$ $\cdots $, $P_{1}(x)$ are polynomials of degree
respectively 17, 19, 20, 21, 22, 23, 24, 23, 22 and 21 given in Appendix E.

With (\ref{defQ}), the differential equation needs 242 unknowns to be found
(counting the polynomial in front of the derivative of order 0, which is
identically null). Our series expansion for $\tilde{\chi}^{(4)}/16w^{16}$,
having only 217 terms in the variable $x=16w^2$, is, thus, not long enough
to let the differential equation be found. This calls for some comments.

The differential equation (\ref{fuchs}) is of {\em minimal order}. It is obvious
that one can obtain other differential equations of {\em greater order}. As
explained in \cite{ze-bo-ha-ma-05} (see Section 4), before the differential
equation built from a series expansion pops out, the singularities computed
as roots of the polynomial in front of the highest derivative reach
stabilized numerical values as the degrees of the polynomials and/or the order
get higher. Even the minimal multiplicity of the singularity can be seen.
This leads us to take the following form for the $\tilde{\chi}^{(4)}$
differential equation
\begin{eqnarray}
\label{ode}
\sum_{i=0}^{q}\,R_{i}(x) \cdot  x^i 
\, (1-x)^i \cdot {\frac{{d^{i}}}{{dx^{i}}}}F(x)\,\,=\,\,\,\,\,0
\end{eqnarray}
where only the physical singularities are explicitly included.
Note that the other (non-apparent) singularity can also be included.
If the differential equation can be identified with the number of series
coefficients at hand, the unnecessary terms will factor out.
A compromise has to be found with respect to how long the series is.
Now, there are many ways to choose the degrees of the
polynomials $R_i(x)$. We have taken
\begin{eqnarray}
\label{degR}
deg \Bigl( R_i(x) \Bigr) = \mu +q -i, \qquad i=0,1, \cdots, q
\end{eqnarray}
in order to have the point at infinity as a {\em regular singular point}\footnote[4]
{This feature can easily be seen, making the change of variable $t=1/x$
and looking for the necessary condition for $t=0$ to be a regular
singular point from which (\ref{degR}) is deduced.}.

If the differential equation of order $q$ and degree $\mu$ exists,
the series of $\tilde{\chi}^{(4)}/16w^{16}$ should have at least $N$ terms, with :
\begin{eqnarray}
N \, = \, \mu  \cdot (q+1) + {{1}\over{2}} \, q \cdot (q+3).
\end{eqnarray}
Staying below the above hyperbola, we have obtained, at  
$q=11$ and $\mu=9$ which requires 185 terms in the series, 
two linearly independent differential equations that satisfy
the remaining 30 terms of the series. The combination of these two
differential equations gives (\ref{fuchs}) which has been checked to be
of minimal order.
Note the fact that besides the Fuchsian differential equation of minimal
order with an apparent polynomial, there are {\em other differential equations
of higher order that require less terms in the series to be identified}.
We plan to report on this feature elsewhere.

The singularities of this differential equation $x=0, 1, 4$ and $x=\infty$
are all regular singular points. The roots of the polynomial $P_{10}(x)$
are {\em apparent singularities}.
One notes, contrary, to $\chi^{(3)}$ that no new singularity is found
besides the known physical and non physical singularities (i.e.,
Nickel's \cite{nickel-99,nickel-00}).
The critical exponents of all these singular
points are given in Table 1 with the maximum power of logarithmic terms
in the solutions. As was the case for $\chi^{(3)}$,
these logarithmic terms appear due to the multiple roots of the
indicial equation.
A noteworthy remark is the occurrence of 
logarithmic terms up to the power 3 for $\chi^{(4)}$ to be compared
with the power 2 for $\chi^{(3)}$ at the singular points $w=\pm 1/4$
and $w=\infty$.

\vskip 0.5cm

\centerline{
\begin{tabular}{|l|l|l|l|}
\hline
&  &  &    \\ 
$x$-singularity & $s$-singularity & Critical exponents in $x$ &  $P$
\\ 
&  &  &    \\ 
\hline
&  &  &    \\
$0$ & $0,\infty$        & $8,3,3,2,2,1,1,0,0,-1/2$          &   $1$ \\ 
&  &  &    \\ 
$1$ & $ \pm 1$              & $3,2,1,1,0,0,0,0,-1,-3/2$       &   $3$ \\ 
&  &  &    \\ 
$4$ & $ \pm {{1}\over{2}} \pm i {{\sqrt{3}}\over{2}}$      & $8,7,13/2,6,5,4,3,2,1,0$     &   $0$ \\ 
&  &  &    \\ 
$\infty $ & $\pm i$     & $5/2, 3/2, 3/2, 1/2, 1/2, 1/2, 1/2, 0, -1/2, -1/2$          &   $3$ \\ 
&  &  &    \\ 
$x_{P}$, 17 roots & $s_{P}$, 68 roots     & $10,8,7,6,5,4,3,2,1,0$      &   $0$ \\ 
&  &  &    \\ \hline
\end{tabular}
}
\vskip 0.2cm
\textbf{Table 1:} Critical exponents for each regular singular point. 
$P$ is the maximum power of the logarithmic terms for each singularity.
$x_{P}$ is any of the 17 roots of $P_{10}(x)$.

\vskip 0.5cm

 It is worth recalling the Fuchsian relation on 
 Fuchsian type equations. Denoting by
 $\, x_1$,  $\, x_2,\,  \cdots \, $,  
$\, x_m$, $\, x_{m+1} \, = \, \infty$, the regular 
singular points of a  Fuchsian type equation
of order $\, q$
and $\, \rho_{j,1}, \, $ $\, \cdots, \, \rho_{j,q}$
  ($j \, = \, 1, \, \cdots\, , m+1$)
the $\, q$ roots of the {\em indicial equation}~\cite{ince-56,Forsyth} corresponding to each 
regular singular point $\, w_j$, the following  Fuchsian relation~\cite{ince-56,Forsyth} 
holds :
\begin{eqnarray}
\label{Fuchs}
\sum_{j\, =\, 1}^{m+1} \,\sum_{k\, =\, 1}^{q} \, \rho_{j,k}\,\,  = \, \,\, 
{{ (m-1)\, \, q \, \, (q-1)} \over {2}}.
\end{eqnarray}
The number of regular singular points here is $\, m+1 \, = \, 21$
corresponding respectively to the 17 roots of $\, P_{10}$,  the
$\, x \, = \, 0, \, 1, \, 4, \, $ regular singular points,
and the point at infinity $\,x \, = \,\infty$.
The Fuchsian relation (\ref{Fuchs}) is actually satisfied here with $\, q\, = 10$,
$\, m\, = \, 20\,$.

Considering the Fuchsian relation (\ref{Fuchs}), but now in the variable $\, s$, 
one first remarks that the $\, \rho_{j,k}$ exponents are {\em the same} as
the ones in the $\, x$ variable for the $\, 68$ roots 
of the apparent polynomial,
as well as for the four $ x=4$ singularities (namely 
$\, s \, = \, \pm {{1}\over{2}} \pm i {{\sqrt{3}}\over{2}}$),
but {\em are multiplied by a factor two} for all the other ones.
This is obvious from the definition of $x=4s^2/(1+s^2)^2$ at $x=0$ and
$x=\infty$, and this can be seen for $x=1$ from :
\begin{eqnarray}
1-x \, = \, \Bigl( {\frac{1-s^2}{1+s^2}} \Bigr)^2
\end{eqnarray}
The Fuchsian relation is actually satisfied in the $\, s$ variable, but now,
 with $\, q\, = 10$, $\, m\, = \, \,77$.

Let us close this section by confirming the dominant singular behavior
at $x=4$ and $x=\infty$ given in \cite{nickel-00}, since this can easily
be done from the critical exponents given in Table 1. Near $x=4$, the
solution behaves as $t^{13/2}$ and near $x=\infty$ as $t^{-1/2}\, \log(t)$,
with respectively $t=4-x$ and $t=1/x$. Fom Table 1, the subdominant
singular behavior at $x=\infty$ is
$t^{3/2}\, \log(t)$ and $t^{1/2}\, \log^k(t)$ (with $k=1,2,3$).

\section{Properties of the Fuchsian differential equation (\ref{fuchs})}
\label{simplsol}
From Table 1, and from the formal solutions around the singular points, it is easy
to find the following simple solutions of (\ref{fuchs}) :
\begin{eqnarray}
{\cal S}_0(x) =\,  constant
\end{eqnarray}
\begin{eqnarray}
{\cal S}_1(x) = \, {\frac{8-12x+3x^2}{8(1-x)^{3/2}}}, 
\end{eqnarray}
\begin{eqnarray}
{\cal S}_2(x) = \, {\frac{2-6x+x^2}{2(1-x)\sqrt{x}}}.
\end{eqnarray}
These solutions correspond to solutions of some differential operators of order one. Let us
call these order one differential operators 
respectively $L_0$, $L_1$ and $L_2$.

\vskip 0.3cm

A remarkable finding {\em is the following solution of
 the Fuchsian differential equation} (\ref{fuchs}) :
\begin{eqnarray}
{\cal S}_3(x) = \,
 {{1}\over{64}} \,x^{2}\cdot {_{2}}F_{1}\Bigl({{5}\over{2}},{{3}\over{2}};3; x \Bigr)
\end{eqnarray}
{\em which is nothing but the two-particle contribution to
 the magnetic susceptibility}, i.e.,
$\tilde{\chi}^{(2)}$ associated with an operator of order two, $N_0$ ($N_0({\cal S}_3) = \,0$).
We will come back to this point later.

The second solution of the order-two operator $N_0$ is given in terms
of the MeijerG function~\cite{Meijer} :
\begin{eqnarray}
\tilde{{\cal S}}_3 (x) =\,{\frac{\pi}{2}}\,  {\rm MeijerG} \left( [[],[1/2,3/2]], [[2,0],[]],x \right)
\end{eqnarray}
which can also be written as
\begin{eqnarray}
\tilde{{\cal S}}_3 (x) = \, {\cal S}_3 (x) \, \log \left( x \right) + \,B (x)
\end{eqnarray}
with : 
\begin{eqnarray}
B(x) = {\frac{1}{12 \, \pi}} \cdot 
\sum _{k=0}^{\infty } {x}^{k}
{\frac{d}{dk}}\, \Bigl(  
{\frac { \Gamma \left( k-1/2 \right) \Gamma \left( k+1/2 \right) } 
{ \Gamma \left( k-1 \right) \Gamma \left( k+1 \right) } }
\Bigr)
\end{eqnarray}

With these five solutions corresponding to three differential operators of order
one, and one differential operator of order two, it is easy to construct 24
factorizations of ${\cal L}_{10}$, the differential operator corresponding to
the Fuchsian differential equation
(\ref{fuchs}), which can be written\footnote[5]{All the operators are such
that the coefficient in front of the highest derivative is $\, +1$.} as :
\begin{eqnarray}
{\cal L}_{10} \, = \,\, O_5  \cdot G(N). 
\end{eqnarray}
$G(N)$ is a shorthand notation of a differential operator of order 5, factorizable
in one operator of order two and three operators of order 1. $G(N)$ has 24
different factorizations involving eight differential operators of order two and
24 operators of order one. The differential operator $O_5$ factorizes
as $M_1 \cdot L_{24}$, i.e., one operator of order four and one operator of
order one. Let us give two examples of the 24 factorizations\footnote[6]{
We denote by $L$ the operators of order 1, by $N$, the operators of order 2
and by $M$ the operators of order 4 (see Appendix F).}:
\begin{eqnarray}
{\cal L}_{10} \, &=& \, M_1  \cdot L_{24} \cdot N_4 \cdot L_{12} \cdot L_3 \cdot L_0 \\
{\cal L}_{10} \, &=& \, M_1  \cdot L_{24} \cdot L_{13} \cdot N_{6} \cdot L_3 \cdot L_0
\end{eqnarray}

This large number of factorizations~\cite{Hoeij,Singer,Dmodule} induces
 the occurrence of intertwiners
\footnote[1]{We thank Jacques-Arthur Weil for usefull comments on the
 equivalence of linear differential operators~\cite{Put} (see equation (5) in~\cite{Dmodule}).}.
In the examples above, one has $N_4 \cdot L_{12}=L_{13} \cdot N_{6}$. Seeking
a similar relation for $L_{24} \cdot N_4$, one finds $N_{9} \cdot L_{25}$.
This last factorization introduces six factorizations that we denote as:
\begin{eqnarray}
{\cal L}_{10} \, = \, \,M_1  \cdot N_{9} \cdot G(L).
\end{eqnarray}
$G(L)$ is a notation for an operator of order four, factorizable as
four operators of order one.

One factorization of $G(L)$ reads:
\begin{eqnarray}
G(L) \, = \,\, L_{25}  \cdot L_{12} \cdot L_3 \cdot L_0.
\end{eqnarray}
This differential operator $G(L)$ that factorizes ${\cal L}_{10}$ at right, obviously has 
 ${\cal S}_0$, ${\cal S}_1$ and ${\cal S}_2$ as
solutions.
The fourth solution (of the order four differential operator $G(L)$) 
can be obtained by order reduction. It reads
\begin{eqnarray}
{\cal S}_4(x) &=& \, {\frac{4\,(x-2)\sqrt{4-x}}{x-1}} +
16 \log{  {\frac{x}{(2+\sqrt{4-x})^2}}   }  \nonumber \\
&&+  16 \, {\cal S}_1(x)  \cdot  \log{g(x)} \, -  
   16\, \sqrt{x} \, {\cal S}_2(x) \, \cdot 
 {_{2}}F_{1}\Bigl({{1}\over{2}},{{1}\over{2}};{{3}\over{2}}; {{x}\over{4}} \Bigr)
\end{eqnarray}
with :
\begin{eqnarray}
g(x) = \,\, {{1}\over{x}} \, \Bigl( (8-9x+2x^2) \, +2 \,(2-x) \cdot \sqrt{(1-x)(4-x)} \Bigr)
\end{eqnarray}

To get more factorizations, we need to obtain  simple solutions 
of ${\cal L}_{10}^{*}$, {\em the adjoint of
 the differential operator} ${\cal L}_{10}$. There is no solution of ${\cal L}_{10}^{*}$
corresponding to an order-one operator, however, {\em we have been  able to 
find a solution corresponding to an operator
of order two} (denoted $N^{*}_8$). This solution\footnote[2]{
We did not look for the second solution of $N^{*}_8$, our purpose being the
factorization of ${\cal L}_{10}$.} of $N^{*}_8$
 is a combination of elliptic integrals
with polynomials of quite large degrees and reads
\begin{eqnarray}
{\cal S}_1^{*}(x) =\,  {\frac{x\, (1-x)^6(4-x)}{3840000 \cdot P_{10}(x)}}
\cdot \Bigl( q_1(x) \,K(x)+q_2(x) \,E(x) \Bigr)
\end{eqnarray}
where
\begin{eqnarray}
K(x)={_{2}}F_{1}\Bigl({{1}\over{2}},{{1}\over{2}};1;x\Bigr), \qquad
E(x)={_{2}}F_{1}\Bigl(-{{1}\over{2}},{{1}\over{2}};1;x\Bigr)
\end{eqnarray}
and :
\begin{eqnarray}
q_1(x) & = &
 (1-x) ( 47352014438400-246257318625280\,x  \nonumber \\
&& +275880211382272\,{x}^{2}+68328139784192\,{x}^{3} \nonumber \\
&& +645943284072448\,{x}^{4}-2774821715853312\,{x}^{5}  \nonumber \\
&& +3217221650489344\,{x}^{6}-683914539437568\,{x}^{7} \nonumber \\
&& -2042467767948624\,{x}^{8}+3083863919521506\,{x}^{9} \nonumber \\
&& -2746206480894969\,{x}^{10}+1558224994851490\,{x}^{11}  \nonumber \\
&& -347705392468761\,{x}^{12}-145625559012638\,{x}^{13}  \nonumber \\
&& +117842186745065\,{x}^{14}-30744722745590\,{x}^{15} \nonumber \\
&& +3089482306025\,{x}^{16}+18651488480\,{x}^{17}    \nonumber \\
&& -21574317760\,{x}^{18}+821760000\,{x}^{19} )
\end{eqnarray}
\begin{eqnarray}
q_2(x) & = &
-47352014438400 +269933325844480\,x  \nonumber \\
&& -390130367987712\,{x}^{2}+27877970018304\,{x}^{3} \nonumber \\
&&  -580571855978496\,{x}^{4} +3135069528473600\,{x}^{5}  \nonumber \\
&& -4488375407386624\,{x}^{6}+1736922901371392\,{x}^{7}  \nonumber \\
&& +2600277912748368\,{x}^{8}-5123144341863018\,{x}^{9} \nonumber \\
&& +5224617790090830\,{x}^{10}-3547418998359865\,{x}^{11} \nonumber \\
&& +1453586336314895\,{x}^{12}-273126255420088\,{x}^{13}    \nonumber \\
&& -8194519962996\,{x}^{14}+11308926014655\,{x}^{15} \nonumber \\
&& -1235672485785\,{x}^{16}-60982101700\,{x}^{17}  \nonumber \\
&& +15765744320\,{x}^{18} -607865600\,{x}^{19}  \nonumber \\
&& -3840000\,{x}^{20}
\end{eqnarray}

This order 2 operator $N_8$ completes the factorization scheme of ${\cal L}_{10}$, the differential
operator of order ten. Let us recap the factorizations
\begin{eqnarray}
\label{factoriz}
{\cal L}_{10}  &=& N_8 \cdot M_{2} \cdot G(L)  \nonumber \\
{\cal L}_{10}  &=& M_1 \cdot N_{9} \cdot G(L)   \nonumber \\
{\cal L}_{10}  &=& M_1 \cdot L_{24} \cdot G(N)   
\end{eqnarray}

$G(L)$ is a differential operator of order four, factorizable in order one operators.
It has six different factorizations involving thirteen differential operators of
order one. $G(N)$ is an operator of order five, factorizable in one operator of
order two and three operators of order one. $G(N)$ has 24 different factorizations
involving eight differential operators of order two, and 24 order one operators
with twelve appearing in $G(L)$.

All these 36 factorizations (\ref{factoriz}) are given in Appendix F.
Appendix F shows that these 36 factorizations can be considered 
as  {\em only one} factorization up to a set of
 equivalence-symmetries~\cite{Put,Dmodule,Hoeij,Weil,Telescoping,Ore,Help}.
\vskip 0.1cm

From the ten solutions of the differential equation (\ref{fuchs}), six
solutions are given explicity, ${\cal S}_0$, ${\cal S}_1$, ${\cal S}_2$,
${\cal S}_4$ (this last being a solution of an order four operator) and
two solutions (${\cal S}_3$, $\tilde{{\cal S}}_3$) corresponding
to the differential equation of $\tilde{\chi}^{(2)}$.
The remaining four solutions are those of the operator of order eight, 
$M_{2} \cdot G(L)$.

From the 36 factorizations shown in Appendix F, those six of the form
$N_8 \cdot M_{2} \cdot G(L)$ are of the most importance. Their occurrence
allows us to get the contribution $\alpha$ of  $\tilde{\chi}^{(2)}$ in the
physical solution  $\tilde{\chi}^{(4)}$ from
$M_{2} \cdot G(L) (\tilde{\chi}^{(4)}-\alpha \tilde{\chi}^{(2)})=0$.
This contribution is obtained easily
and gives
\begin{eqnarray}
\label{contrchi2}
\tilde{\chi}^{(4)} \,=\, {{1}\over{3}} \tilde{\chi}^{(2)}+
\Phi_4 \, 
\end{eqnarray}
where $\Phi_4$ is solution of the order eight
 differential operator $M_{2} \cdot G(L)$.

\vskip 0.3cm

Recall that the same situation occurred for the differential equation of
order seven for $\tilde{\chi}^{(3)}$.
 One can see that the rational solution $\, w/(1-4\, w)$
 occurring\footnote[3]{This corresponds
to the solution $S_1$ given in \cite{ze-bo-ha-ma-04,ze-bo-ha-ma-05}
which is $S_1=\tilde{\chi}^{(1)}/2$.}
 in the differential equation of
order seven for $\tilde{\chi}^{(3)}$
is nothing but $\tilde{\chi}^{(1)}$.
With this remark we can rewrite a decomposition
of $\tilde{\chi}^{(3)}$ we gave in~\cite{ze-bo-ha-ma-04,ze-bo-ha-ma-05} as follows :
\begin{eqnarray}
\label{contrchi1}
\tilde{\chi}^{(3)} \,=\, {{1}\over{6}} \tilde{\chi}^{(1)}+
\Phi_3.
\end{eqnarray}
where $\Phi_3$ is solution of the order six differential
 operator (noted $L_6$ in \cite{ze-bo-ha-ma-04,ze-bo-ha-ma-05}).

\section{Comments and speculations}
\label{conjec}
Denoting $\, {\bf L}_4$ the order ten differential operator
 associated with the ordinary differential
equation satisfied by $\tilde{\chi}^{(4)}$, and
more generally, $\, {\bf L}_n$ the differential operators
 associated with the ordinary differential
equation satisfied by $\tilde{\chi}^{(n)}$, one has :
\begin{eqnarray}
\label{L4Ki4}
 {\bf L}_4 (\tilde{\chi}^{(4)})
\,=\,\, {\bf L}_4 (\tilde{\chi}^{(2)})\,=\,\,  0.
\end{eqnarray}
Furthermore, one can see that the rational solution $\, w/(1-4\, w)$, occurring
 in the differential equation of
order seven for $\tilde{\chi}^{(3)}$,
is nothing but $\tilde{\chi}^{(1)}$ and thus one has :
\begin{eqnarray}
\label{L3Ki3}
 {\bf L}_3 (\tilde{\chi}^{(3)})\,=\, {\bf L}_3 (\tilde{\chi}^{(1)})\,=\,\, 0.
\end{eqnarray}
Both relations come from the factorizations of the differential operators
corresponding to $\tilde{\chi}^{(3)}$ and $\tilde{\chi}^{(4)}$.

At this point, it is tempting to make some conjectures generalizing relations
(\ref{L4Ki4}, \ref{L3Ki3}) and relations (\ref{contrchi2},\ref{contrchi1}).

From (\ref{L4Ki4}) and (\ref{L3Ki3}) the conjecture is
\begin{eqnarray}
&& {\bf L}_{2n+1} (\tilde{\chi}^{(2n+1)})\,=\,
{\bf L}_{2n+1} (\tilde{\chi}^{(1)}) \,=\,\, 0, \nonumber \\
&& {\bf L}_{2n} (\tilde{\chi}^{(2n)})\,=\,
{\bf L}_{2n} (\tilde{\chi}^{(2)}) \,=\,\, 0
\end{eqnarray}
meaning that the differential operator of $\chi^{(1)}$ (respectively $\chi^{(2)}$)
right divides the differential operator corresponding to $\chi^{(2n+1)}$
(respectively $\chi^{(2n)}$).

One stronger conjecture is to expect the same situation as in
(\ref{contrchi2},\ref{contrchi1}) occurring in the higher particle
contributions, i.e., 
\begin{eqnarray}
\tilde{\chi}^{(2n)} \,&=&\, \alpha_{2n} \, \tilde{\chi}^{(2)}+
\Phi_{2n},  \nonumber \\
\tilde{\chi}^{(2n+1)} \,&=&\, \alpha_{2n+1} \,  \tilde{\chi}^{(1)}+
\Phi_{2n+1}  
\end{eqnarray}
where $\Phi_{2n}$ (respectively $\Phi_{2n+1}$) is solution of a differential
operator that right divides the differential operator ${\bf L}_{2n}$
(respectively ${\bf L}_{2n+1}$) and is not divisible by the differential
operator ${\bf L}_{2}$ (respectively ${\bf L}_{1}$).
In this situation, it is easy to obtain the numbers $\alpha_n$ which give
the contribution of $\tilde{\chi}^{(1)}$ and $\tilde{\chi}^{(2)}$ in the
higher $\tilde{\chi}^{(n)}$'s as explained for $\tilde{\chi}^{(4)}$ (see
text before (\ref{contrchi2})).

Let us note that both conjectures are free from any constraint due to the
singularities that occur in the differential equations, since
any $\tilde{\chi}^{(2n+1)}$ (respectively $\tilde{\chi}^{(2n)}$) has
the singularities occurring in $\tilde{\chi}^{(1)}$
(respectively $\tilde{\chi}^{(2)}$).

A much stronger conjecture is to expect any $\chi^{(m)}$ to be "embedded"
in any $\chi^{(n)}$ where $m$ divides $n$ with same odd-even parity.
\begin{eqnarray}
\label{LnKin}
&& {\bf L}_n (\tilde{\chi}^{(n)})\,=\, {\bf L}_n (\tilde{\chi}^{(m)}) \,=\,\, 0, \nonumber \\
&&{\bf L}_n \,=\,\, {\bf L}_n^{(m)} \cdot {\bf L}_{m}.
\end{eqnarray}
This means that $\, {\bf L}_n$ might be built from the {\em least common left multiple}
(lclm) of the differential operators associated with the
 $\, {\bf L}_m$'s where $\, m$ divides the integer $\, n$ respecting
 even-odd parity.
The fact that $n$ should be multiple of $m$ is due to the non physical
singularities appearing in the differential equations. For instance,
no  "embedding" like $\chi^{(3)}$ being solution of the differential
equations of $\chi^{(5)}$ has to be expected.
 Whether  $\chi^{(3)}$ is a solution of the
differential equation satisfied by, e.g., $\chi^{(9)}$ is not ruled out,
the (non-apparent) singularities of the first being (non-apparent) 
singularities of the last.
But, then, the new singularities discovered \cite{ze-bo-ha-ma-04} for
$\chi^{(3)}$ have to occur in the differential equation of $\chi^{(9)}$.

Similarly one can expect, with this last conjecture, the following relations for  
$\, {\bf L}_{12}$ and $\, {\bf L}_{15}$ (associated with $\, \tilde{\chi}^{(12)}$
 and $\, \tilde{\chi}^{(15)}$) 
\begin{eqnarray}
\label{L15Ki15}
&& {\bf L}_{12} (\tilde{\chi}^{(12)})
\,=\,\, {\bf L}_{12} (\tilde{\chi}^{(6)})\,=\,\,
{\bf L}_{12} (\tilde{\chi}^{(4)})\,=\,\,{\bf L}_{12} (\tilde{\chi}^{(2)})
\,=\,\, 0 \nonumber \\
&&{\bf L}_{15} (\tilde{\chi}^{(15)})
\,=\,\, {\bf L}_{15} (\tilde{\chi}^{(5)})\,=\,\,
{\bf L}_{15} (\tilde{\chi}^{(3)})\,=\,\,{\bf L}_{15} (\tilde{\chi}^{(1)})
\,=\,\, 0
\end{eqnarray}
and thus :
\begin{eqnarray}
\label{L15Ki15bis}
&&{\bf L}_{12}\,=\,\,{\bf L}_{12}^{(6)} \cdot {\bf L}_{6}\,=\,\,{\bf L}_{12}^{(4)} \cdot {\bf L}_{4}
\,=\,\,{\bf L}_{12}^{(2)} \cdot {\bf L}_{2}  \nonumber \\
&&{\bf L}_{15}\,=\,\,{\bf L}_{15}^{(5)} \cdot {\bf L}_{5}\,=\,\,{\bf L}_{15}^{(3)} \cdot {\bf L}_{3}
\,=\,\,{\bf L}_{15}^{(1)} \cdot {\bf L}_{1}.
\end{eqnarray}

For instance, finding the differential operator satisfied by $\, \tilde{\chi}^{(15)}$
would enable us to see if the differential operator $\, {\bf L}_{15}$
actually has some relation with the least common left multiple
of $\, {\bf L}_1$, $\, {\bf L}_3$ and $\, {\bf L}_5$.

\section{Conclusion}
\label{conclu}
Considering the isotropic Ising square lattice model
susceptibility, we extended our ``expansion method'' (that allowed us to
find the differential equation satisfied by $\chi^{(3)}$), to the   
four-particle
contribution to the susceptibility, namely $\chi^{(4)}$. We first obtained
$\chi^{(4)}$ as a {\em fully integrated multisum on products of four hypergeometric
functions}, and obtained a long series for $\chi^{(4)}$.
 From this long series, we gave the Fuchsian differential equation
of order ten satisfied by $\chi^{(4)}$.
This differential equation has a rich structure in terms of factorizations.

We have given in closed form six,
of the ten solutions of the differential equation (\ref{fuchs}),
namely, ${\cal S}_0$, ${\cal S}_1$, ${\cal S}_2$,
${\cal S}_4$, ${\cal S}_3$ and $\tilde{{\cal S}}_3$.
The remaining four solutions are those of an operator of order eight.

One of these solutions is highly remarkable: {\em it is
 actually the two-particle contribution}
$\chi^{(2)}$. 
{\em A similar situation also occured for the differential equation of the three particle
contribution} $\chi^{(3)}$, {\em which actually had} $\chi^{(1)}$ {\em as solution}.

In general, for all the  $\tilde{\chi}^{(n)}$'s,  it is tempting to expect 
 $\tilde{\chi}^{(1)}$
to be a solution of the differential equation satisfied by the $\tilde{\chi}^{(2 n+1)}$' s and
 $\tilde{\chi}^{(2)}$ to be a solution of the differential equation satisfied by
the  $\tilde{\chi}^{(2 n)}$'s.

Beyond, one can contemplate much stronger conjectures corresponding to 
further "embedding",
 namely  $\tilde{\chi}^{(m)}$ being solution of the differential
equation of $\tilde{\chi}^{(n)}$, when $\, n$ is a multiple
 of $\, m$, $\, n$ and  $\, m$ having the same parity.

A confirmation of these various embeddings and conjectures is crucial
because it corresponds to a {\em global structure} of the 
$\tilde{\chi}^{( n)}$'s  {\em and thus of} $\chi$  {\em itself}.
We will investigate this feature in a future publication.

\hskip 2cm

\textbf{Acknowledgments} We thank Jacques-Arthur Weil for valuable
comments on factorizations of differential operators.
We would like to thank A. J. Guttmann and J. Dethridge for
 extensive calculations on our series. We
would like to thank particularly J. Dethridge for rewriting our formal
programs into 
a masterpiece C++ program that confirmed our series and our final ODE result 
up to order 250.    We thank the medicis formal calculation platform, at 
the stix laboratory of the Ecole Polytechnique, for enabling us to
perform some formal calculations on their 8 Giga-memory machine. 
(S. B) and (S. H) acknowledge partial support from PNR3-Algeria.
(N. Z.) is thankful to T. Hamidouche and H. Mazrou for letting him
use their computer facilities to generate the first 130 terms of the series.

\section{Appendix A}
\label{appena}
In this Appendix, we show that, with $H^{(4)}$ taken as in (\ref{defH4}),
i.e., not the alternative simplified form we use in this paper,
our expansion method applied to $\tilde{\chi}^{(4)}$ produces {\em five} summations
on products of {\em four} hypergeometric functions.
Note that this method is applicable to any $\chi^{(n)}$.

Let us write the product $R^{(4)}\cdot H^{(4)}$ as
\begin{equation}
\label{defR4H4}
R^{(4)}\cdot H^{(4)}=T^{(4)}\cdot A^{(4)}  
\end{equation}
with :
\begin{equation}
T^{(4)}=\, R^{(4)}\cdot \prod_{i<j}\frac{\tilde{x}_{i}\tilde{x}_{j}}
{(1-\tilde{x}_{i}\tilde{x}_{j})^{2}}
\label{defT4}
\end{equation}
\begin{equation}
\label{defA4}
A^{(4)}=\frac{1}{4!}\cdot \prod_{i<j}(Z_{i}-Z_{j})^{2}.  
\end{equation}

By standard expansion of the quantity $T^{(4)}$, defined by (\ref{defT4}),
in the variables $\tilde{x}_{i}$, one obtains
\begin{eqnarray}
\label{calc-T41}
T^{(4)} &=&\sum_{p=0}^{\infty }\sum_{i_{1}=0}^{\infty
}\sum_{i_{2}=0}^{\infty }\sum_{i_{3}=0}^{\infty }\sum_{i_{4}=0}^{\infty
}\sum_{i_{5}=0}^{\infty }\sum_{i_{6}=0}^{\infty }\left( 1+\theta
(p-1)\right) \cdot \prod_{k=1}^{6}(i_{k}+1)\cdot  \nonumber \\
&&\tilde{x}_{1}^{p+i_{1}+i_{2}+i_{3}+3}\tilde{x}_{2}^{p+i_{1}+i_{4}+i_{5}+3}
\tilde{x}_{3}^{p+i_{2}+i_{4}+i_{6}+3}\tilde{x}_{4}^{p+i_{3}+i_{5}+i_{6}+3}
\end{eqnarray}
where $\theta (x)$ is the step function given in (\ref{defTHETA}). Defining
\begin{eqnarray}
\label{defni}
n_{1} &=&p+i_{1}+i_{2}+i_{3}, \qquad n_{2} =p+i_{1}+i_{4}+i_{5}  \nonumber \\
n_{3} &=&p+i_{2}+i_{4}+i_{6}, \qquad n_{4} =p+i_{3}+i_{5}+i_{6},   
\end{eqnarray}
one can solve (\ref{defni}) in the indices $(i_{3},i_{4},i_{5},i_{6})$, to obtain:
\begin{eqnarray}
\label{defi3}
i_{3} &=& -\left( p+i_{1}+i_{2} \right) +n_{1}  \\
\label{defi4}
i_{4} &=& -\left( p+i_{1}+i_{2} \right) +{{1}\over{2}} \left(n_{1}+n_{2}+n_{3}-n_{4} \right)
   \\
\label{defi5}
i_{5} &=& i_{2} -{{1}\over{2}} \left( n_{1}-n_{2}+n_{3}-n_{4} \right)    \\
\label{defi6}
i_{6} &=& i_{1} -{{1}\over{2}} \left( n_{1}+n_{2}-n_{3}-n_{4} \right)
\end{eqnarray}
All these indices should be {\em integers}, inducing constraints on the $n_i$'s
and limitations on the summations of the remaining indices
(i.e., $p$, $i_1$, $i_2$). With this change of summation indices,
$T^{(4)}$ becomes
\begin{eqnarray}
\label{EXP-T4}
T^{(4)} &=& \sum_{n_{1}=0}^{\infty }\sum_{n_{2}=0}^{\infty}
\sum_{n_{3}=0}^{\infty }\sum_{n_{4}=0}^{\infty}\, \nonumber \\
&&\quad \quad  C(n_{1},n_{2},n_{3},n_{4}) \cdot
 \tilde{x}_{1}^{n_{1}+3}\tilde{x}_{2}^{n_{2}+3}
\tilde{x}_{3}^{n_{3}+3}\tilde{x}_{4}^{n_{4}+3}
\end{eqnarray}
and, since $T^{(4)}$ is completely symmetric in the variables 
$\tilde{x}_{i}$, the coefficient $C(n_{1},n_{2},n_{3},n_{4})$ is completely
symmetric in the $n_{i}$ indices and is given by summation on the remaining
indices  $p$, $i_1$ and $i_2$:
\begin{eqnarray}
\label{calc-C1}
C(n_{1},n_{2},n_{3},n_{4}) &=&\sum_{p=0}^{\infty }\sum_{i_{1}=0}^{\infty}
\sum_{i_{2}=0}^{\infty }\left( 1+\theta (p-1)\right) \cdot
\prod_{k=1}^{6}(i_{k}+1) \nonumber \\
&& \times \theta (i_{3})\cdot \theta(i_{4})\cdot \theta (i_{5})\cdot \theta (i_{6})
 \cdot
 \sigma (n_{1}+n_{2}+n_{3}+n_{4})
\nonumber \\
&&  \times \theta \Bigl( {{1}\over{2}} \left(n_{1}+n_{2}+n_{3}-n_{4} \right) \Bigr)
\end{eqnarray}
where theta
functions (\ref{defTHETA}) of indices $(i_{3},i_{4},i_{5},i_{6})$ 
take place  to keep track of the fact that these indices 
should be positive integers. The
symbol $\sigma (n)$ defined as:
\begin{equation}
\label{defsig}
\sigma (n)={{1}\over{2}} \left( 1+(-1)^{n} \right)  
\end{equation}
comes from the fact that $i_{3},i_{4},i_{5},i_{6}$ are
integers, due to the right  most terms at right-hand-side of (\ref{defi4}-\ref{defi6}),
the $n_i$'s should verify $n_{1}+n_{2}+n_{3}+n_{4}={\rm even\,\,integer}$.
The argument in the last theta in (\ref{calc-C1}) comes from (\ref{defi4}).

Let us define the index $q$ and its upper limit of summation
(see (\ref{defi3},\ref{defi4})):
\begin{eqnarray}
\label{defq}
q &=&p+i_{1}+i_{2}   \\
\label{defq0}
q_{0} &=&\min (n_{1},\frac{n_{1}+n_{2}+n_{3}-n_{4}}{2})  
\end{eqnarray}

Furthermore, the coefficient $C(n_{1},n_{2},n_{3},n_{4})$ being symmetric in all its
arguments, it is sufficient to compute it in the case where:
\begin{equation}
\label{confni}
n_{1} \leq  n_{2} \leq  n_{3} \leq  n_{4}  
\end{equation}
The coefficient $C(n_{1},n_{2},n_{3},n_{4})$ now becomes:
\begin{eqnarray}
\label{calc-C3}
C(n_{1},n_{2},n_{3},n_{4}) &=&\theta (\frac{n_{1}+n_{2}+n_{3}-n_{4}}{2}%
)\cdot \sigma (n_{1}+n_{2}+n_{3}+n_{4})  \nonumber \\
&&\sum_{q=0}^{q_{0}}\sum_{p=0}^{q} \left( \left( 1+\theta (p-1)\right) \cdot 
\sum_{i_{1}=0}^{q-p} \, ( \prod_{k=1}^{6}(i_{k}+1)) \right) 
\end{eqnarray}
where the indices $i_{3},i_{4},i_{5},i_{6}$ are given by
(\ref{defi3}-\ref{defi6})
and $i_{2}=q-p-i_{1}$. The three summations can be performed and one obtains
\begin{eqnarray}
\label{calc-C4}
C(n_{1},n_{2},n_{3},n_{4}) &=&\, \theta (\frac{n_{1}+n_{2}+n_{3}-n_{4}}{2}
)\cdot \sigma (n_{1}+n_{2}+n_{3}+n_{4})  \nonumber \\
&& \cdot {q_0+4 \choose 4}Q(n_{1},n_{2},n_{3},n_{4},q_{0})   
\end{eqnarray}
with the polynomial $Q(n_{1},n_{2},n_{3},n_{4},q_{0})$ given by
\begin{eqnarray}
\label{defPOL}
&& Q(n_{1},n_{2},n_{3},n_{4},q_{0}) \, =\, \frac{1}{8}
Q_{0}(n_{1},n_{2},n_{3},n_{4})+\frac{q_{0}}{2520}
Q_{1}(n_{1},n_{2},n_{3},n_{4})  \nonumber \\
&& \quad \qquad +\frac{q_{0}^{2}}{7560}Q_{2}(n_{1},n_{2},n_{3},n_{4})+
\frac{q_{0}^{3}}{1890}Q_{3}(n_{1},n_{2},n_{3},n_{4})  \nonumber \\
&& \quad \qquad +\frac{q_{0}^{4}}{540}(92-81n_{1}-9n_{2}-9n_{3}+63n_{4})+
\frac{4q_{0}^{5}}{135}  
\end{eqnarray}
with :
\begin{eqnarray}
\label{def-P0}
&& Q_{0}(n_{1},n_{2},n_{3},n_{4}) \, =\, (1+n_{1})(n_{1}+n_{2}-n_{3}-n_{4}-2) 
\nonumber \\
&&\quad \qquad (n_{1}-n_{2}+n_{3}-n_{4}-2)(2+n_{1}+n_{2}+n_{3}-n_{4}) 
\end{eqnarray}
\begin{eqnarray}
\label{def-P1}
&& Q_{1}(n_{1},n_{2},n_{3},n_{4}) \, = \,
-2216+2472n_{1}+1194n_{1}^{2}-1281n_{1}^{3}+126n_{1}^{4}
\nonumber \\
&&\qquad -48n_{2}+1392n_{1}n_{2}-987n_{1}^{2}n_{2}+126n_{1}^{3}n_{2}+774n_{2}^{2}+441n_{1}n_{2}^{2}
\nonumber \\
&&\qquad -126n_{1}^{2}n_{2}^{2}+147n_{2}^{3}-126n_{1}n_{2}^{3}-48n_{3}+1392n_{1}n_{3}-987n_{1}^{2}n_{3}
\nonumber \\
&&\qquad +126n_{1}^{3}n_{3}-1548n_{2}n_{3}-882n_{1}n_{2}n_{3}+252n_{1}^{2}n_{2}n_{3}-147n_{2}^{2}n_{3} 
\nonumber \\
&&\qquad +126n_{1}n_{2}^{2}n_{3}+774n_{3}^{2}+441n_{1}n_{3}^{2}-126n_{1}^{2}n_{3}^{2}-147n_{2}n_{3}^{2}
\nonumber \\
&&\qquad +126n_{1}n_{2}n_{3}^{2}+147n_{3}^{3}-126n_{1}n_{3}^{3}-2568n_{4}-708n_{1}n_{4}
\nonumber \\
&&\qquad +2709n_{1}^{2}n_{4}-378n_{1}^{3}n_{4}-288n_{2}n_{4}+1134n_{1}n_{2}n_{4}-252n_{1}^{2}n_{2}n_{4}
\nonumber \\
&&\qquad -147n_{2}^{2}n_{4}+126n_{1}n_{2}^{2}n_{4}-288n_{3}n_{4}+1134n_{1}n_{3}n_{4}-147n_{3}^{2}n_{4}
\nonumber \\
&&\qquad +294n_{2}n_{3}n_{4}-252n_{1}n_{2}n_{3}n_{4}-252n_{1}^{2}n_{3}n_{4}+126n_{1}n_{2}n_{4}^{2}
\nonumber \\
&&\qquad +126n_{1}n_{3}^{2}n_{4}-486n_{4}^{2}-1575n_{1}n_{4}^{2}+378n_{1}^{2}n_{4}^{2}-147n_{2}n_{4}^{2}
\nonumber \\
&&\qquad -147n_{3}n_{4}^{2}+126n_{1}n_{3}n_{4}^{2}+147n_{4}^{3}-126n_{1}n_{4}^{3}
\end{eqnarray}
\begin{eqnarray}
\label{def-P2}
&& Q_{2}(n_{1},n_{2},n_{3},n_{4}) \, =\,
-2992-4446n_{1}+6534n_{1}^{2}-1449n_{1}^{3}-1926n_{2}
\nonumber \\
&&\qquad -819n_{1}^{2}n_{2}-270n_{2}^{2}+945n_{1}n_{2}^{2}+315n_{2}^{3}-1926n_{3}+2808n_{1}n_{3}
\nonumber \\
&&\qquad -819n_{1}^{2}n_{3}+540n_{2}n_{3}-1890n_{1}n_{2}n_{3}-315n_{2}^{2}n_{3}+2808n_{1}n_{2} 
\nonumber \\
&&\qquad -270n_{3}^{2}+945n_{1}n_{3}^{2}-315n_{2}n_{3}^{2}+315n_{3}^{3}+594n_{4}-315n_{3}n_{4}^{2} 
\nonumber \\
&&\qquad -8532n_{1}n_{4}+3213n_{1}^{2}n_{4}-1728n_{2}n_{4}+1134n_{1}n_{2}n_{4}+315n_{4}^{3} 
\nonumber \\
&&\qquad -315n_{2}^{2}n_{4}-1728n_{3}n_{4}+1134n_{1}n_{3}n_{4}+630n_{2}n_{3}n_{4} 
\nonumber \\
&&\qquad -315n_{3}^{2}n_{4}+1998n_{4}^{2}-2079n_{1}n_{4}^{2}-315n_{2}n_{4}^{2}
\end{eqnarray}
\begin{eqnarray}
\label{def-P3}
&& Q_{3}(n_{1},n_{2},n_{3},n_{4}) \, = \,
520-1215n_{1}+495n_{1}^{2}-207n_{2}+144n_{1}n_{2}
\nonumber \\
&&\qquad +144n_{1}n_{3}+270n_{2}n_{3}-135n_{3}^{2}+801n_{4}-738n_{1}n_{4}
\nonumber \\
&&\qquad -108n_{3}n_{4}+243n_{4}^{2}-135n_{2}^{2}-207n_{3}-108n_{2}n_{4}  
\end{eqnarray}

From the definition of $\tilde{\chi}^{(4)}$ in (\ref{def KHI4Tilda}), and from
(\ref{defR4H4},\ref{EXP-T4}), one gets :
\begin{eqnarray}
\label{KHI4T-form1}
\tilde{\chi}^{(4)} &=& 16\, w^{16} \cdot \sum_{n_{1}=0}^{\infty}
\sum_{n_{2}=0}^{\infty }\sum_{n_{3}=0}^{\infty }\sum_{n_{4}=0}^{\infty}
w^{n_{1}+n_{2}+n_{3}+n_{4}} \times   \\
&& \qquad C(n_{1},n_{2},n_{3},n_{4})\cdot M(n_{1},n_{2},n_{3},n_{4})  \nonumber
\end{eqnarray}
The expansion method on the variables $x_i$ introduces some summations
free of any angular dependence. These summations appear in the coefficient
$C(n_{1},n_{2},n_{3},n_{4})$ which is a kind of
``geometrical factor'' that appears for all the $\chi^{(n)}$'s with $n \ge 4$.

At this step the integrations have not yet been performed. They are contained in:
\begin{eqnarray}
\label{defM}
M(n_{1},n_{2},n_{3},n_{4}) &=&
\int_{0}^{2\pi }\frac{d\phi _{1}}{2\pi } \int_{0}^{2\pi }\frac{d\phi _{2}}{2\pi }
\int_{0}^{2\pi }\frac{d\phi _{3}}{2\pi } \nonumber \\
&&\quad \quad \quad \quad \quad \quad 
\left( \prod_{i=1}^{4}y_{i}\cdot x_{i}^{n_{i}+3}\right) \cdot A^{(4)}
\end{eqnarray}

Due to its structure, it is obvious that $M(n_{1},n_{2},n_{3},n_{4})$\ is
completely symmetric in the indices $n_{i}$. We write 
$M(n_{1},n_{2},n_{3},n_{4})$ as an integral over the four
angles $(\phi _{1},\phi _{2},\phi _{3},\phi _{4})$, by
introducing Dirac delta function
$\delta (\phi _{1}+\phi _{2}+\phi _{3}+\phi _{4})$ as a Fourier expansion
that reads:
\begin{equation}
\label{delta}
2\, \pi \, \delta (\phi _{1}+\phi _{2}+\phi _{3}+\phi
_{4})=\sum_{k=-\infty }^{\infty }(Z_{1}Z_{2}Z_{3}Z_{4})^{k}  
\end{equation}
$M(n_{1},n_{2},n_{3},n_{4})$ thus becomes 
\begin{equation}
\label{CalculM}
M(n_{1},n_{2},n_{3},n_{4})=\sum_{k=-\infty }^{\infty
}M(n_{1},n_{2},n_{3},n_{4};k)  
\end{equation}
where :
\begin{eqnarray}
\label{CalculMk}
M(n_{1},n_{2},n_{3},n_{4};k) &=&\int_{0}^{2\pi }\frac{d\phi _{1}}{2\pi }%
\int_{0}^{2\pi }\frac{d\phi _{2}}{2\pi }\,\int_{0}^{2\pi }\frac{d\phi _{3}}{%
2\pi }\int_{0}^{2\pi }\frac{d\phi _{4}}{2\pi }\, \nonumber \\
&&  \quad  \left( \prod_{i=1}^{4}y_{i}\cdot x_{i}^{n_{i}+3}\right)  
 A^{(4)}\cdot (Z_{1}Z_{2}Z_{3}Z_{4})^{k}.  
\end{eqnarray}

The Fourier expansion (\ref{Four-yxn}) implies the following integration rule
\begin{equation}
\int_{0}^{2\pi }\frac{d\phi }{2\pi }\, (yx^{m})\cdot Z^{j}\, =\, \, A(m,j).
\label{integrule}
\end{equation}
The calculation of (\ref{CalculMk}) thus becomes straightforward and 
\textit{does not induce any summation}. The quantity $M(n_{1},n_{2},n_{3},n_{4};k)$
comes out as a huge expression with 201 terms, each of them being a
product of four hypergeometric functions. Due to the definition (\ref{CalculM})
where $k$ runs from $-\infty$ to $\infty$, and to the symmetry of
(\ref{CalculM}) under the permutation
of the $n_i$'s, these 201 terms reduce to only 16 terms and
$M(n_{1},n_{2},n_{3},n_{4};k)$ reads:
\begin{eqnarray}
&& M(n_1,n_2,n_3,n_4;k) \, \equiv \, 
    A(n_1,k) A(n_2,k) A(n_3,k) A(n_4,k)    \nonumber \\
&&\qquad -A(n_1,k) A(n_2,k) A(n_3,k) A(n_4,k+4)  \nonumber \\
&&\qquad A(n_1,k) A(n_2,k) A(n_3,k+2) A(n_4,k+2)  \nonumber \\
&&\qquad  +2 A(n_1,k) A(n_2,k) A(n_3,k+1) A(n_4,k+3)   \nonumber \\
&&\qquad  -3 A(n_1,k) A(n_2,k+1) A(n_3,k+1) A(n_4,k+2)  \nonumber \\
&&\qquad  -A(n_1,k) A(n_2,k) A(n_3,k+3) A(n_4,k+5)      \nonumber \\
&&\qquad  +A(n_1,k) A(n_2,k) A(n_3,k+4) A(n_4,k+4)      \nonumber \\
&&\qquad  -2A(n_1,k) A(n_2,k+1) A(n_3,k+3) A(n_4,k+4)      \nonumber \\
&&\qquad  +2 A(n_1,k) A(n_2,k+1) A(n_3,k+2) A(n_4,k+5)  \nonumber \\
&&\qquad  +2 A(n_1,k) A(n_2,k+2) A(n_3,k+3) A(n_4,k+3)  \nonumber \\
&&\qquad  -A(n_1,k) A(n_2,k+2) A(n_3,k+2) A(n_4,k+4)      \nonumber \\
&&\qquad  +A(n_1,k) A(n_2,k+2) A(n_3,k+4) A(n_4,k+6)    \nonumber \\
&&\qquad  -A(n_1,k) A(n_2,k+2) A(n_3,k+5) A(n_4,k+5)      \nonumber \\
&&\qquad  -A(n_1,k) A(n_2,k+3) A(n_3,k+3) A(n_4,k+6)    \nonumber \\
&&\qquad  +2 A(n_1,k) A(n_2,k+3) A(n_3,k+4) A(n_4,k+5)  \nonumber \\
&&\qquad  -A(n_1,k) A(n_2,k+4) A(n_3,k+4) A(n_4,k+4)
\end{eqnarray}
which is equal to (\ref{CalculMk}) for summation purposes.

Finally, collecting (\ref{KHI4T-form1}, \ref{CalculM}), $\tilde{\chi}^{(4)}$
can be written as :
\begin{eqnarray}
\label{KHI4T-form2}
\tilde{\chi}^{(4)} &=&16\, w^{16}\cdot \sum_{n_{1}=0}^{\infty}
\sum_{n_{2}=0}^{\infty }\sum_{n_{3}=0}^{\infty }\sum_{n_{4}=0}^{\infty}
w^{n_{1}+n_{2}+n_{3}+n_{4}} \times  \nonumber \\
&&\qquad\quad \quad C(n_{1},n_{2},n_{3},n_{4})\cdot M(n_{1},n_{2},n_{3},n_{4})
\end{eqnarray}

\section{Appendix B}
\label{appenb} 
In this Appendix, we give the explicit derivation of the
expression (\ref{H4 simple}) given in Section 2.

Considering $H^{(4)}$ as in (\ref{H4form1}), we define the
following relations
\begin{equation}
\label{PX}
\prod_{i<j}\frac{\tilde{x}_{i}-\tilde{x}_{j}}{1-\tilde{x}_{i}\tilde{x}_{j}}
=P_{12-34}^{(x)}-\left(1\leftrightarrow 3\right) -\left( 2\leftrightarrow 3\right) 
\end{equation}
\begin{equation}
\label{PZ}
\prod_{i<j}\frac{Z_{i}-Z_{j}}{1-Z_{i}Z_{j}}=P_{12-34}^{(Z)}-\left(
1\leftrightarrow 3\right) -\left( 2\leftrightarrow 3\right) 
\end{equation}
where $P_{12-34}^{(x)}$ is given in (\ref{PX12-34}).  Using the
constraint (\ref{contrainte}), $P_{12-34}^{(Z)}$ reads :
\begin{eqnarray}
\label{PZ-12-34}
&&\,\, P_{12-34}^{(Z)} =\, 
 \tilde{A}^{(4)} + (Z_{2}-Z_{1})(Z_{3}-Z_{4}) (1+Z_{1}Z_{2}+Z_{3}Z_{4}) \\
\label{whatA}
&&{\rm with} \quad \quad \quad  \widehat{A}^{(4)} =\, 
(Z_{2}-Z_{1})(Z_{3}-Z_{4})\cdot
\frac{Z_{1}Z_{2}}{\left( 1-Z_{1}Z_{2}\right) ^{2}}
\end{eqnarray}
Note that, while $H^{(4)}$ and $R^{(4)}$\ are completely symmetric under the
permutation of the angles $\phi _{i}$, the right hand side of (\ref{PX},\ref{PZ})
are anti-symmetric under the same transformation.

From (\ref{H4form1}, \ref{PX}, \ref{PZ}) and using the symmetry of
the rest of the integrand (namely $R^{(4)}$\ times the product over $y_{i}$
variables), we can write $H^{(4)}$ as:
\begin{equation}
\label{H4-form2}
H^{(4)}\, \equiv \, \, \frac{1}{8}P_{12-34}^{(Z)}\cdot \left( P_{12-34}^{(x)}-
\left( 1 \leftrightarrow 3 \right) -\left( 2 \leftrightarrow 3 \right) \right)
\end{equation}

We obtain $H^{(4)}$, given in (\ref{H4 simple}), from (\ref{H4-form2}) with
the first term at the right hand side of (\ref{PZ-12-34}).
Again using the property of complete symmetry of the integrand,
 that part of (\ref{H4-form2}) with the second term at the
 right hand side of (\ref{PZ-12-34})
 can be written as:
\begin{eqnarray}
&& -{\frac{1}{8}} \, P_{12-34}^{(x)}\cdot
\Bigl( (Z_1-Z_2)(Z_3-Z_4)(1+Z_1Z_2+Z_3Z_4) \nonumber \\
&& -\left( 1 \leftrightarrow 3 \right) -\left( 2 \leftrightarrow 3 \right)
\Bigr)
\end{eqnarray}
and {\em is identically null}.

\section{Appendix C}
\label{appenc} 
In this Appendix, we give the explicit integration rules
corresponding to the form (\ref{formeintegrand}) given in the main text.
For this purpose, let us consider 
\begin{eqnarray}
\label{Jdef}
J(n_{1},n_{2},n_{3},n_{4}) &=&
\int_{0}^{2\pi }\frac{d\phi _{1}}{2\pi }
\int_{0}^{2\pi }\frac{d\phi _{2}}{2\pi }
\int_{0}^{2\pi }\frac{d\phi _{3}}{2\pi } \nonumber \\
&& \qquad \quad \quad \left( \prod_{i=1}^{4}y_{i}\, x_{i}^{n_{i}}\right) \cdot \widehat{A}^{(4)}  
\end{eqnarray}
where $\widehat{A}^{(4)}$ is given in (\ref{whatA}).
Expression $J(n_{1},n_{2},n_{3},n_{4})$ has the following properties:
\begin{eqnarray}
\label{Jprops}
J(n,n,n_{3},n_{4}) &=&\,J(n_{1},n_{2},n,n)\,=\,\, 0  \nonumber \\
J(n_{1},n_{2},n_{3},n_{4}) &=& -J(n_{2},n_{1},n_{3},n_{4})\, =\,\, J(n_{3},n_{4},n_{1},n_{2}) 
\end{eqnarray}
where the last identity comes from the fact that
\begin{eqnarray}
\frac{Z_{1}Z_{2}}{(1-Z_{1}Z_{2})^{2}}\, =\,\, \,
 \frac{Z_{3}Z_{4}}{(1-Z_{3}Z_{4})^{2}}.
\end{eqnarray}
Changing the integration variables from $(\phi _{1},\phi _{2},\phi _{3})$
to $(\phi =\phi _{1}+\phi _{2},\phi _{2},\phi _{3})$, $\widehat{A}^{(4)}$,
defined in (\ref{whatA}), becomes:
\begin{equation}
\label{A4Tilda-formZ}
\widehat{A}^{(4)}=\frac{1}{\left( 1-Z\right) ^{2}}\cdot \left[
  \left( \frac{1}{Z_{2}Z_{3}}
+Z_{2}Z_{3}\right)\cdot Z
 -\frac{Z_{2}}
{Z_{3}}\, -\frac{Z_{3}}{Z_{2}} \, Z^{2}
 \right]   
\end{equation}
with $Z=Z_{1}Z_{2}$. Taking $\widehat{A}^{(4)}$ in the form
(\ref{A4Tilda-formZ}), expanding $y_{1}x_{1}^{n_{1}}$\ and $y_{4}x_{4}^{n_{4}}$,
 and using the Fourier expansion (\ref{Four-yxn}), the integration over $\phi _{2}
$ and $\phi _{3}$ is straightforward\footnote[4]{
see the integration rule (\ref{integrule}).}. Some standard manipulations give:
\begin{eqnarray}
J(n_{1},n_{2},n_{3},n_{4}) &=& -\int_{0}^{2\pi }\frac{d\phi }{2\pi }\cdot
\sum_{i=-\infty }^{\infty }\sum_{j=-\infty }^{\infty } \nonumber \\
&& \quad \quad D(n_{1},n_{2};i)\, D(n_{3},n_{4};j) \cdot 
\frac{Z^{j-i+1}}{\left( 1-Z\right) ^{2}}
\label{J-form1}
\end{eqnarray}
where :
\begin{equation}
\label{Dn1n2k-def}
D(n_{1},n_{2};k)=\, A(n_{1},k+1)\, A(n_{2},k)-A(n_{1},k)\, 
A(n_{2},k+1). 
\end{equation}
From the definition (\ref{def-A}) of $A(n,k)$ , one can easily show that
\begin{equation}
\label{D-props}
D(n_{1},n_{2};-k)=-D(n_{1},n_{2};k-1)  
\end{equation}
With this property (\ref{D-props}),
and after some manipulations, $J(n_{1},n_{2},n_{3},n_{4})$ reads
\begin{equation}
\label{J-form2}
J(n_{1},n_{2},n_{3},n_{4})=\, \sum_{i=0}^{\infty }\sum_{j=0}^{\infty
}D(n_{1},n_{2};i)\, D(n_{3},n_{4};j)\, N(i,j)  
\end{equation}
where
\begin{equation}
\label{Kij-def}
N(i,j)=\int_{0}^{2\pi }\frac{d\phi }{2\pi }\, Z^{-j-i}\, \frac{
1-Z^{2i+1}}{1-Z}\, \frac{1-Z^{2j+1}}{1-Z}  
\end{equation}
which, by expansion in powers of $Z$ and integration, simply reads:
\begin{equation}
\label{Kij-expression}
N(i,j) \,=\, 2\min (i,j)+1.  
\end{equation}
Finally, taking the form (\ref{J-form2}), together with (\ref{Kij-expression}),
and using the definition (\ref{def-A}) of $A(n,k)$, we obtain:
\begin{eqnarray}
\label{J-final-form}
&& J(n_{1},n_{2},n_{3},n_{4}) \, = \,  {{1}\over{2}} \sum_{j=0}^{\infty }
\sum_{k=0}^{\infty}
w^{2j+4k+2}\, \left( 2k+1\right) \,
\left( 1+\theta \left(j-1\right) \right) \cdot   \nonumber \\
&&\quad  \Bigl(
d(n_{1},n_{2};k)\, d(n_{3},n_{4};j+k)+d(n_{1},n_{2};j+k)\,d(n_{3},n_{4};k)
\Bigr)
\end{eqnarray}
where $d(n_{1},n_{2};k)$ is defined in (\ref{def-d}).

\section{Appendix D}
\label{append}
In this Appendix, we give all the necessary details of the
derivation of the result (\ref{KHI4T-FINAL}) given in the text. We consider $%
\tilde{\chi}^{(4)}$ in the form (\ref{def KHI4Tilda}), with $H^{(4)}$ given
in (\ref{H4 simple}).

The first step is to consider the expansion of the product
$R^{(4)} \cdot P_{12-34}^{(x)}$.
Let us give the expansions, in the $x_{i}$'s variables, for the first two
terms in the brackets in (\ref{PX12-34}):
\begin{eqnarray}
\label{xexp-1}
R^{(4)} \cdot \frac{\tilde{x}_{1}\tilde{x}_{2}}{1-\tilde{x}_{1}\tilde{x}_{2}}
\frac{\tilde{x}_{3}\tilde{x}_{4}}{1-\tilde{x}_{3}\tilde{x}_{4}}
&=&\sum_{m=0}^{\infty }\sum_{l=0}^{\infty }\left( \tilde{x}_{1}\tilde{x}_{2}\right)
^{m}\left( \tilde{x}_{3}\tilde{x}_{4}\right) ^{l}\cdot   \\
&&\theta (m-1)\, \theta (l-1)\, \left( 2\min (m,l)-1\right) \nonumber
\end{eqnarray}
\begin{eqnarray}
\label{xexp-2}
R^{(4)} \cdot \frac{\left( \tilde{x}_{1}\tilde{x}_{2}\right) ^{2}}{1-\tilde{x}_{1}\tilde{x}_{2}}
&=&\sum_{m=0}^{\infty }\sum_{l=0}^{\infty }
\left( \tilde{x}_{1}\tilde{x}_{2}\right) ^{m}\left(\tilde{x}_{3}\tilde{x}_{4}\right)^{l}
 \cdot  \\
&&\theta (m-l-2)\, \theta (m-2)\, \left( 1+\theta (l-1)\right). \nonumber
\end{eqnarray}
$R^{(4)} \cdot P_{12-34}^{(x)}$ can be expanded as
\begin{eqnarray}
\label{xexp-R4P12-34}
R^{(4)}\cdot P_{12-34}^{(x)} &=& (x_{1}-x_{2})\cdot (x_{3}-x_{4})\cdot \sum_{m=0}^{\infty
}\sum_{l=0}^{\infty }w^{2m+2l+2} \cdot   \\
&& \left( x_{1}x_{2}\right) ^{m}\left(x_{3}x_{4}\right) ^{l} \cdot c(m,l)  \nonumber
\end{eqnarray}
with :
\begin{eqnarray}
\label{cml-def}
c(m,l) &=&\theta (m-1)\, \theta (l-1) \cdot \left( 2\min
(m,l)-1\right)   \nonumber \\
&&+\theta (m-l-2)\, \theta (m-2) \cdot \left( 1+\theta (l-1)\right)  
\nonumber \\
&&+\theta (l-m-2)\, \theta (l-2) \cdot \left( 1+\theta (m-1)\right). 
\end{eqnarray}
Considering $\tilde{\chi}^{(4)}$ in the form (\ref{def KHI4Tilda}), with $%
H^{(4)}$ given by (\ref{H4 simple}), taking the expansion of
$R^{(4)}\cdot P_{ij-kl}^{(x)}$ in the form (\ref{xexp-R4P12-34}), performing the angular
integration using the definition (\ref{Jdef}), and collecting
all terms, gives
\begin{equation}
\label{KHI4-form1}
\tilde{\chi}^{(4)}=\,\, 16\,w^{6} \cdot \sum_{m=0}^{\infty }\sum_{p=0}^{\infty }
\, {{1}\over{2}} w^{2m+2p} \, c(m,p) \, I(m,p)  
\end{equation}
where :
\begin{eqnarray}
\label{Iml-def}
I(m,p) &=&\, J(m,m+1,p,p+1)-J(m,p,m+1,p+1) \nonumber \\
&& \quad \quad  -J(m,p+1,p,m+1).
\end{eqnarray}
Using the identities (\ref{Jprops}) for expression
$J(n_{1},n_{2},n_{3},n_{4})$, one has :
\begin{eqnarray}
\label{Iml-props}
I(p,m) &=&I(m,p)  \nonumber \\
I(m,m) &=&I(m,m+1)=\, 0.  
\end{eqnarray}
The symmetry of $c(m,p)$,  together with the relations (\ref{Iml-props}),
allow us, after some manipulations, to write $\tilde{\chi}^{(4)}$ as
\begin{equation}
\label{KHI4-form2}
\tilde{\chi}^{(4)}=\, 16\, w^{10}\cdot \sum_{m=0}^{\infty }\sum_{n=0}^{\infty }w^{4m+2n}%
\, \left( 2m+1\right) \cdot I(m,n+m+2)  
\end{equation}
where we have used the following identity $c(m,n+m+2)=2m+1$.

From the previous equation, taking the definition (\ref{Iml-def}) together
with the expression (\ref{J-final-form}), one obtains
\begin{eqnarray}
\label{KHI4-form3}
\tilde{\chi}^{(4)} &=& 16w^{12}\cdot \sum_{m,k,n,j=0}^{\infty }w^{4m+4k+2n+2j}\, 
\left( 2m+1\right) \, \left( 2k+1\right)\times  \nonumber \\
&& {{1}\over{2}} \, \Bigl( 1+\theta \left(j-1\right) \Bigr)\cdot V(m,k,n,j-2)  
\end{eqnarray}
where $V(m,k,n,j)$ is defined in (\ref{defHYP-KHI4}), with the constraints :
\begin{equation}
\label{Kmknj-props}
V(m,k,n,-2) \, = \, V(m,k,n,-1) \, =\, 0  
\end{equation}
Thus, $\tilde{\chi}^{(4)}$ reads :
\begin{eqnarray}
\label{KHI4-prefinal form}
\tilde{\chi}^{(4)} &=& 16\, w^{16} \cdot \sum_{m,k,n,j=0}^{\infty }w^{4m+4k+2n+2j} \times \nonumber \\
&& \qquad\quad \quad \left( 2m+1\right) \cdot \left( 2k+1\right) \cdot V(m,k,n,j).
\end{eqnarray}
The last simplification comes from the following identity on $V(m,k,n,j)$
\begin{equation}
\label{Fmknj-props}
V(k,m,j,n)=\, V(m,k,n,j).  
\end{equation}
which is obtained with the help of the symmetry of $a(n,k)$, namely
$a(n,k)=a(k,n)$.

Finally, using the general identity
\begin{eqnarray}
\label{general identity}
\sum_{p_{1}=0}^{\infty }\sum_{p_{2}=0}^{\infty}s(p_{1},p_{2}) &=&
\sum_{p=0}^{\infty }\sum_{i=0}^{\infty }\left( 1+\theta(i-1)\right) \times \nonumber \\
&&\quad \quad  {{1}\over{2}} \left( s(p+i,p)+s(p,p+i) \right)
\end{eqnarray}
and the identity (\ref{Fmknj-props}), the expression (\ref{KHI4-prefinal
form}) for $\tilde{\chi}^{(4)}$ can, after some manipulation, be written in
the form (\ref{KHI4T-FINAL}) displayed in the text.

\section{Appendix E}
\begin{eqnarray}
P_{10}(x) &=&
   192598769664000-943722860380160\,x \nonumber \\
&& +1154055263764480\,{x}^{2}+223612469147648\,{x}^{3} \nonumber \\
&& +498965092419008\,{x}^{4}-4709824359388336\,{x}^{5} \nonumber \\
&& +6098813329440179\,{x}^{6}-2687506699337617\,{x}^{7} \nonumber \\
&& -752969324018818\,{x}^{8}+1919011581320339\,{x}^{9} \nonumber \\
&& -1526656430056013\,{x}^{10}+660280621356073\,{x}^{11} \nonumber \\
&& -134468923815612\,{x}^{12}+7980003107181\,{x}^{13}  \nonumber \\
&& +671991155376\,{x}^{14}-17122807680\,{x}^{15} \nonumber \\
&& -5111390208\,{x}^{16}+43929600\,{x}^{17}
\end{eqnarray}

\begin{eqnarray}
P_9(x) &=&
 39290149011456000-297447362118287360\,x   \nonumber \\
&& +785045342889902080\,{x}^{2}-738039510467911680\,{x}^{3}     \nonumber \\
&& +115430038068948224\,{x}^{4}-1014554971757285568\,{x}^{5}    \nonumber \\
&& +3530827584228091380\,{x}^{6}-4191676869770939784\,{x}^{7}   \nonumber \\
&& +2038314404276922993\,{x}^{8}+302242744021242129\,{x}^{9}    \nonumber \\
&& -1228406406705470056\,{x}^{10}+1043856642401535101\,{x}^{11} \nonumber \\
&& -506114460611418219\,{x}^{12}+139123852417110463\,{x}^{13}   \nonumber \\
&& -18756086615529738\,{x}^{14}+585977956065027\,{x}^{15}       \nonumber \\
&& +80289868813296\,{x}^{16}+338076239232\,{x}^{17} \nonumber \\
&& -554882826240\,{x}^{18}+4524748800\,{x}^{19}
\end{eqnarray}

\begin{eqnarray}
P_8(x) &=&
  -197221140135936000+2237384185085952000\,x  \nonumber \\
&& -9353251554131968000\,{x}^{2}+ 17417661939469352960\,{x}^{3} \nonumber \\
&& -12764493479941255168\,{x}^{4}+ 5721656503410538176\,{x}^{5} \nonumber \\
&& -31513487740861948240\,{x}^{6}+ 74327124872294722259\,{x}^{7} \nonumber \\
&& -74793556916527926194\,{x}^{8}+ 31202576844037093699\,{x}^{9} \nonumber \\
&& +8393285642528898691\,{x}^{10}- 21583664946799604164\,{x}^{11} \nonumber \\
&& +16831944005767359920\,{x}^{12}- 7706219890490209273\,{x}^{13} \nonumber \\
&& +2036277808796847523\,{x}^{14}- 267301496692115881\,{x}^{15} \nonumber \\
&& +8265260736796220\,{x}^{16}+ 1095242827496832\,{x}^{17}  \nonumber \\
&& +4127794059264\,{x}^{18}-7320844222464\,{x}^{19} \nonumber \\
&& +58514227200\,{x}^{20}
\end{eqnarray}

\begin{eqnarray}
&& P_7(x) =
   764231918026752000-16971026878221516800\,x \nonumber \\
&&\quad  + 115553546628494786560\,{x}^{2}-357128015448284200960\,{x}^{3}   \nonumber \\
&&\quad  +527925844930277396480\,{x}^{4}-327328761204753779200\,{x}^{5}     \nonumber \\
&&\quad  +264654998297650108784\,{x}^{6}-1129192842855079731378\,{x}^{7}    \nonumber \\
&&\quad  +2154212025640162544967\,{x}^{8}-1923633718567453943246\,{x}^{9}   \nonumber \\
&&\quad  +708245350538757849333\,{x}^{10}+257335760183269925607\,{x}^{11}   \nonumber \\
&&\quad  -530877971609916541978\,{x}^{12}+387828430484350338476\,{x}^{13}   \nonumber \\
&&\quad  -169278351316199852707\,{x}^{14}+43213759197299807177\,{x}^{15}    \nonumber \\
&&\quad  -5540993509581072863\,{x}^{16}+171042063471920404\,{x}^{17}        \nonumber \\
&&\quad  +21596554335396864\,{x}^{18}+67975771296768\,{x}^{19}              \nonumber \\
&&\quad  -139524297965568\,{x}^{20}+ 1091386982400\,{x}^{21}
\end{eqnarray}

\begin{eqnarray}
&& P_6(x) =
   2366653681631232000+15657511469396787200\,x \nonumber \\
&&\quad   -418375361108283228160\,{x}^{2}+2413740972333717913600\,{x}^{3}    \nonumber \\
&&\quad  -6304389540169208659968\,{x}^{4}+7885479774610742167552\,{x}^{5}    \nonumber \\
&&\quad  -4194762998349204963264\,{x}^{6}+4787870532372577669360\,{x}^{7}    \nonumber \\
&&\quad  -18347979159424299125323\,{x}^{8}+30953188609834801509491\,{x}^{9}  \nonumber \\
&&\quad  -25453291295484088365791\,{x}^{10}+8618062554065795648668\,{x}^{11} \nonumber \\
&&\quad  +3501102053489313155231\,{x}^{12}-6497943932928673199052\,{x}^{13}  \nonumber \\
&&\quad  +4524991429929970477637\,{x}^{14}-1899642804937798765560\,{x}^{15}  \nonumber \\
&&\quad  +470903276066137118684\,{x}^{16}-59187343449427131707\,{x}^{17}     \nonumber \\
&&\quad  +1841781629480835082\,{x}^{18}+217686233501214240\,{x}^{19}         \nonumber \\
&&\quad  +505738155562752\,{x}^{20}-1358525988175872\,{x}^{21}               \nonumber \\
&&\quad  +10379070873600\,{x}^{22}
\end{eqnarray}

\begin{eqnarray}
&& P_5(x) =
  -4240254512922624000+62906863287743283200\,x             \nonumber \\
&&\quad   -176844924247909335040\,{x}^{2}-926585736920055152640\,{x}^{3}         \nonumber \\
&&\quad  +6926618061828909694976\,{x}^{4}-17270267632607743877120\,{x}^{5}        \nonumber \\
&&\quad  +19088006038191340885888\,{x}^{6}-7801708756272263419344\,{x}^{7}        \nonumber \\
&&\quad  +10929800840921163293912\,{x}^{8}-45123693732451251946951\,{x}^{9}        \nonumber \\
&&\quad  +72469347127122031920503\,{x}^{10}-57004940985415997264335\,{x}^{11}        \nonumber \\
&&\quad  +18768946563950160696028\,{x}^{12}+7071447265880540507399\,{x}^{13}         \nonumber \\
&&\quad  -12955437249853164400164\,{x}^{14}+8753288431922263351485\,{x}^{15}   \nonumber \\
&&\quad  -3567263867581256107712\,{x}^{16}+863585677517038259840\,{x}^{17}     \nonumber \\
&&\quad  -106841634606075117511\,{x}^{18}+3388641803935920834\,{x}^{19}        \nonumber \\
&&\quad  +366870695477680992\,{x}^{20}+460940531245824\,{x}^{21}               \nonumber \\
&&\quad  -2210964576190464\,{x}^{22}+16454358835200\,{x}^{23}
\end{eqnarray}

\begin{eqnarray}
&& P_4(x) =
   2366653681631232000 - 121147207913255731200\,x  \nonumber \\
&&\quad +1204116891510820044800\,{x}^{2} - 4471702378853957632000\,{x}^{3}    \nonumber \\
&&\quad  +3882247113210548060160\,{x}^{4} + 19050981728126452531200\,{x}^{5}   \nonumber \\
&&\quad  -60173463653707794252800\,{x}^{6} + 61647224332324323946560\,{x}^{7}  \nonumber \\
&&\quad  -8902468613107108100784\,{x}^{8} + 12826049043414461768619\,{x}^{9}   \nonumber \\
&&\,\,\,  -130168614999998943881548 {x}^{10} + 224001261613376678757254 {x}^{11}  \nonumber \\
&&\quad  -177856326190823713976989 {x}^{12} + 61054390293590247070213 {x}^{13}   \nonumber \\
&&\quad  +16702537875405165273835\,{x}^{14} - 34504858377234313267137\,{x}^{15}    \nonumber \\
&&\quad  +23145136640956770024901\,{x}^{16} - 9263503411249994637068\,{x}^{17}    \nonumber \\
&&\quad  +2206094527520924501173\,{x}^{18} -270104745702122623849\,{x}^{19}        \nonumber \\
&&\quad  +8842661536198397940\,{x}^{20} +853061197515225600\,{x}^{21}        \nonumber \\
&&\quad  -104681638038528\,{x}^{22}  -4965596931735552\,{x}^{23}              \nonumber \\
&&\quad  +35863949721600\,{x}^{24}
\end{eqnarray}

\begin{eqnarray}
&& P_3(x) =
   -11833268408156160000+296345375978029056000\,x      \nonumber \\
&&\quad   - 2144074027349363916800\,{x}^{2}   +6754783276113459937280\,{x}^{3}     \nonumber \\
&&\quad  -8346802888116933918720\,{x}^{4}   -3858851854352328468480\,{x}^{5}       \nonumber \\
&&\quad  +18632235498956809831680\,{x}^{6}   -5896832882414934369872\,{x}^{7}     \nonumber \\
&&\quad  -20092581953702558202578\,{x}^{8}   +9039258514198720200789\,{x}^{9}     \nonumber \\
&&\quad  +30407815233273202124059\,{x}^{10}   -44070504829347073747333\,{x}^{11} \nonumber \\
&&\quad  +24353838992694268159678\,{x}^{12}   -2634376094447814881673\,{x}^{13}    \nonumber \\
&&\quad  -5786419996453628346078\,{x}^{14}   +5219893433655665719185\,{x}^{15}      \nonumber \\
&&\quad  -2420429903132324343712\,{x}^{16}   +639931684854726532004\,{x}^{17}       \nonumber \\
&&\quad  -85290635341359290409\,{x}^{18}   +3288947218141266164\,{x}^{19}           \nonumber \\
&&\quad  +249618983188012416\,{x}^{20}   -1896484503705600\,{x}^{21}              \nonumber \\
&&\quad  -1406458574438400\,{x}^{22}-9870981120000\,{x}^{23}
\end{eqnarray}

\begin{eqnarray}
&& P_2(x) =  36978963775488000000-561176784419998924800\,x    \nonumber \\
&&\quad  + 3011344824723628359680\,{x}^{2}- 7448695810925636485120\,{x}^{3}     \nonumber \\
&&\quad  + 7522952469866279936000\,{x}^{4} + 1571640457415428633600\,{x}^{5}    \nonumber \\
&&\quad   - 6448286163781893470272\,{x}^{6} - 5723224826762698171248\,{x}^{7}    \nonumber \\
&&\quad   +17523982043162827496339\,{x}^{8} - 10871737924438446047496\,{x}^{9}   \nonumber \\
&&\quad   - 1585720216853837551908\,{x}^{10} + 4808127638071252147363\,{x}^{11}  \nonumber \\
&&\quad   - 1996433111317397547653\,{x}^{12} - 255305750980018902593\,{x}^{13}   \nonumber \\
&&\quad   + 746533058733318361745\,{x}^{14} - 451215245481551846087\,{x}^{15}    \nonumber \\
&&\quad   + 139082951228390470494\,{x}^{16} - 20614123445854835699\,{x}^{17}     \nonumber \\
&&\quad   + 931298896466844519\,{x}^{18}   + 53593802936851536\,{x}^{19}        \nonumber \\
&&\quad   - 845315534083200\,{x}^{20}-297344753664000\,{x}^{21}       \nonumber \\
&&\quad   + 2006484480000\,{x}^{22}
\end{eqnarray}

\begin{eqnarray}
&& P_1(x) =
   - 14878398029915750400   +132065832050507120640\,x      \nonumber \\
&&\quad   - 489750894279960821760\,{x}^{2}   +843938312965142528000\,{x}^{3}    \nonumber \\
&&\quad   - 436158709746751846400\,{x}^{4}   -504050378774640931456\,{x}^{5}    \nonumber \\
&&\quad   + 362705915662042967696\,{x}^{6}   +930077072222989626372\,{x}^{7}    \nonumber \\
&&\quad   - 1516628759565419342933\,{x}^{8}   +919766159750634600991\,{x}^{9}  \nonumber \\
&&\quad   - 236614768621801865201\,{x}^{10}   -4425339208940491244\,{x}^{11}    \nonumber \\
&&\quad   + 15751985529081115811\,{x}^{12}   +356035582683101460\,{x}^{13}       \nonumber \\
&&\quad   - 3725421245175199951\,{x}^{14}   +1704676309420947212\,{x}^{15}       \nonumber \\
&&\quad   - 304427228661108302\,{x}^{16}   +16257997137256137\,{x}^{17}        \nonumber \\
&&\quad   + 635239438099728\,{x}^{18}   -16870300732800\,{x}^{19}            \nonumber \\
&&\quad   - 3641326080000\,{x}^{20}  + 23063040000\,{x}^{21}
\end{eqnarray}

\section{Appendix F}
\label{appendf}
Let us recall a few elementary facts on factorization
of differential operators, without 
introducing any sophisticated ${\cal D}$-module theory. 
Two differential operators $\, {\cal D}_1$ and ${\cal D}_2$ of order $\, N$ are 
said to be equivalent (see for instance equation (5) in~\cite{Dmodule})
 if there exist two (intertwining) differential operators
$\, R$ and $\,S$ of orders at most $\, N-1$, such that :
\begin{eqnarray}
\label{intertwin}
{\cal D}_2 \cdot R \, = \, \, S \cdot {\cal D}_1 .
\end{eqnarray}
The factorization
of differential operators is not in general unique. There 
can be several factorizations, {\em even an infinite number} 
of factorizations, as can be seen in the simple exemple :
\begin{eqnarray}
\label{infinite}
 {{ d^2} \over {dx^2}} \, = \, \,  \Bigl( {{ d} \over {dx}} 
\, + {{a} \over {a\, x\, +b }} \Bigr)
\cdot \Bigl( {{d} \over {dx}} \, - {{a} \over {a\, x\, +b }} \Bigr)
\end{eqnarray}
where the factorization (\ref{infinite}) holds for every constant $\, a$ and $\, b$.
This can be seen as a consequence of the relation :
\begin{eqnarray}
\label{infinite2}
\Bigl( {{ d} \over {dx}} \, - {{a} \over {a\, x\, +b}} \Bigr) \cdot
 \Bigl (a\, x\, + b\Bigr) 
\, = \, \, \Bigl( a\, x\, +b \Bigr) \cdot {{ d} \over {dx}}
\end{eqnarray}
which means that the multiplication by $\,a\, x\, +b$ transforms 
the  solutions of the differential
operator $\, d/dx $ into a solution of  $\, d/dx\, -a/(a\, x\, +b)$.  
Relation (\ref{infinite2}) is a simple exemple of 
equivalence (\ref{intertwin}) of two differential operators. 
The occurrence of this infinite set of factorizations
can be seen to be a consequence of the fact that the 
various differential operators  of  (\ref{infinite}), (\ref{infinite2})
have solutions the ratio of which can be  rational functions. In general 
this is not the case and one will only have a {\em finite number} of factorizations,
these various factorizations being basically the same {\em up 
to some permutation of the differential operators}, 
each operator in the factorization being changed (``by some transmutation'' (\ref{intertwin}))
{\em into an equivalent one} during the permutation process.

This is typically the situation we encounter with the 
operator ${\cal L}_{10}$ associated with the ordinary differential
equation of order ten satisfied by $\tilde{\chi}^{(4)}$.

This order ten operator ${\cal L}_{10}$  has 36
factorizations with two operators of order four ($M_1$ and $M_2$), ten operators
of order two ($N_0, N_1, \cdots, N_9$) and 26 operators of order one
($L_0, L_1, \cdots L_{25}$). {\em All these operators have
 been normalized in such a way that the coefficient
of the highest derivative is normalized to $\, +1$.}
The operators $L_0$, $L_1$, $L_2$, $N_0$ and $N_8$ below are such that
$L_0({\cal S}_0)=0$, $L_1({\cal S}_1)=0$, $L_2({\cal S}_2)=0$, $N_0({\cal S}_3)=0$ and
$N_8^{*}({\cal S}_1^{*})=0$. These  36
factorizations\footnote[7]{
The DFactor command~\cite{DEtools} of  DEtools can give the 
second factorization in (\ref{firstset}),
namely ${\cal L}_{10} \, = \, \,$ $M_1 \cdot N_9 \cdot L_{25} \cdot L_{12} \cdot L_3 \cdot L_0$.}
of  ${\cal L}_{10}$ read
\begin{eqnarray}
\label{firstset}
{\cal L}_{10} &=& N_8 \cdot M_{2} \cdot L_{25} \cdot L_{12} \cdot L_3 \cdot L_0  \nonumber \\
 &=& M_1 \cdot N_{9} \cdot L_{25} \cdot L_{12} \cdot L_3 \cdot L_0  \nonumber \\
 &=& M_1 \cdot L_{24} \cdot N_4 \cdot L_{12} \cdot L_3 \cdot L_0   \nonumber \\
 &=& M_1 \cdot L_{24} \cdot L_{13} \cdot N_6 \cdot L_3 \cdot L_0   \nonumber \\
 &=& M_1 \cdot L_{24} \cdot L_{13} \cdot L_{17} \cdot N_3 \cdot L_0  \nonumber \\
 &=& M_1 \cdot L_{24} \cdot L_{13} \cdot L_{17} \cdot L_{11} \cdot N_0  
\end{eqnarray}

\begin{eqnarray}
{\cal L}_{10} &=& N_8 \cdot M_{2} \cdot L_{25} \cdot L_{14} \cdot L_4 \cdot L_0  \nonumber \\
 &=& M_1 \cdot N_{9} \cdot L_{25} \cdot L_{14} \cdot L_4 \cdot L_0   \nonumber \\
 &=& M_1 \cdot L_{24} \cdot N_4 \cdot L_{14} \cdot L_4 \cdot L_0   \nonumber \\
 &=& M_1 \cdot L_{24} \cdot L_{15} \cdot N_7 \cdot L_4 \cdot L_0   \nonumber \\
 &=& M_1 \cdot L_{24} \cdot L_{15} \cdot L_{16} \cdot N_3 \cdot L_0  \nonumber \\
 &=& M_1 \cdot L_{24} \cdot L_{15} \cdot L_{16} \cdot L_{11} \cdot N_0  
\end{eqnarray}

\begin{eqnarray}
{\cal L}_{10} &=& N_8 \cdot M_{2} \cdot L_{25} \cdot L_{18} \cdot L_5 \cdot L_1  \nonumber  \\
 &=& M_1 \cdot N_{9} \cdot L_{25} \cdot L_{18} \cdot L_5 \cdot L_1    \nonumber  \\
 &=& M_1 \cdot L_{24} \cdot N_4 \cdot L_{18} \cdot L_5 \cdot L_1    \nonumber  \\
 &=& M_1 \cdot L_{24} \cdot L_{19} \cdot N_5 \cdot L_5 \cdot L_1    \nonumber  \\
 &=& M_1 \cdot L_{24} \cdot L_{19} \cdot L_{23} \cdot N_1 \cdot L_1  \nonumber  \\
 &=& M_1 \cdot L_{24} \cdot L_{19} \cdot L_{23} \cdot L_{10} \cdot N_0  
\end{eqnarray}

\begin{eqnarray}
{\cal L}_{10} &=& N_8 \cdot M_{2} \cdot L_{25} \cdot L_{14} \cdot L_6 \cdot L_1  \nonumber  \\
 &=& M_1 \cdot N_{9} \cdot L_{25} \cdot L_{14} \cdot L_6 \cdot L_1    \nonumber  \\
 &=& M_1 \cdot L_{24} \cdot N_4 \cdot L_{14} \cdot L_6 \cdot L_1    \nonumber  \\
 &=& M_1 \cdot L_{24} \cdot L_{15} \cdot N_7 \cdot L_6 \cdot L_1    \nonumber  \\
 &=& M_1 \cdot L_{24} \cdot L_{15} \cdot L_{20} \cdot N_1 \cdot L_1  \nonumber  \\
 &=& M_1 \cdot L_{24} \cdot L_{15} \cdot L_{20} \cdot L_{10} \cdot N_0  
\end{eqnarray}

\begin{eqnarray}
{\cal L}_{10} &=& N_8 \cdot M_{2} \cdot L_{25} \cdot L_{18} \cdot L_7 \cdot L_2  \nonumber  \\
 &=& M_1 \cdot N_{9} \cdot L_{25} \cdot L_{18} \cdot L_7 \cdot L_2    \nonumber  \\
 &=& M_1 \cdot L_{24} \cdot N_4 \cdot L_{18} \cdot L_7 \cdot L_2    \nonumber  \\
 &=& M_1 \cdot L_{24} \cdot L_{19} \cdot N_{5} \cdot L_7 \cdot L_2  \nonumber  \\
 &=& M_1 \cdot L_{24} \cdot L_{19} \cdot L_{21} \cdot N_2 \cdot L_2  \nonumber  \\
 &=& M_1 \cdot L_{24} \cdot L_{19} \cdot L_{21} \cdot L_9 \cdot N_0  
\end{eqnarray}

\begin{eqnarray}
{\cal L}_{10} &=& N_8 \cdot M_{2} \cdot L_{25} \cdot L_{12} \cdot L_8 \cdot L_2  \nonumber  \\ 
 &=& M_1 \cdot N_{9} \cdot L_{25} \cdot L_{12} \cdot L_8 \cdot L_2    \nonumber  \\ 
 &=& M_1 \cdot L_{24} \cdot N_4 \cdot L_{12} \cdot L_8 \cdot L_2    \nonumber  \\ 
 &=& M_1 \cdot L_{24} \cdot L_{13} \cdot N_{6} \cdot L_8 \cdot L_2  \nonumber  \\ 
 &=& M_1 \cdot L_{24} \cdot L_{13} \cdot L_{22} \cdot N_2 \cdot L_2  \nonumber  \\ 
 &=& M_1 \cdot L_{24} \cdot L_{13} \cdot L_{22} \cdot L_9 \cdot N_0  
\end{eqnarray}

All these 36 factorizations can be seen as the consequence of ``transmutations'' 
like (\ref{intertwin}), where one differential operator becomes an equivalent one, 
 thus yielding some elementary permutation of the operator.  
For instance, the two differential operators of order four, namely 
$M_1$ and  $M_2$, are equivalent, their ``intertwiners'' being two 
order two differential operators $N_8$ and $N_9$ :
\begin{eqnarray}
\label{NMNM}
N_8 \cdot M_2 \,=\,\,  M_1 \cdot N_9
\end{eqnarray}
Actually, one sees clearly that the two first factorizations in (\ref{firstset}) 
are just deduced  from one another by (\ref{NMNM}). 

Nine order-two  differential operators $N_0, \cdots, N_8$
can be seen to be related by various equivalence relations
requiring the introduction of 26 differential operators $L_i$'s
of order one (the intertwiners) :
\begin{eqnarray}
N_1 \cdot L_1 &=& L_{10} \cdot N_0, \qquad  N_2 \cdot L_2 = L_9 \cdot N_0 \nonumber   \\
N_3 \cdot L_0 &=& L_{11} \cdot N_0, \qquad  N_4 \cdot L_{12} = L_{13} \cdot N_6 \nonumber  \\
N_4 \cdot L_{14} &=& L_{15} \cdot N_7, \qquad N_4 \cdot L_{18} = L_{19} \cdot N_5  \\
N_5 \cdot L_5 &=& L_{23} \cdot N_1, \qquad N_5 \cdot L_7 = L_{21} \cdot N_2 \nonumber   \\
N_6 \cdot L_3 &=& L_{17} \cdot N_3, \qquad N_6 \cdot L_8 = L_{22} \cdot N_2 \nonumber  \\
N_7 \cdot L_4 &=& L_{16} \cdot N_3, \qquad N_7 \cdot L_6 = L_{20} \cdot N_1\nonumber   \\
N_9 \cdot L_{25} &=& L_{24} \cdot N_4. \nonumber  
\end{eqnarray}

The 26 order-one differential operators $L_i$'s 
can also be seen to be related by various equivalence relations.
The $L_i$'s being order one differential operators, their intertwiners
($R$ and $S$ in (\ref{intertwin})) should be order zero differential operators, that is
some function $\, f_{i,j}$. As a consequence of the normalization to $\, +1$ of the highest order
derivative (here $\, d/dx$), the two intertwiners $R$ and $S$ in (\ref{intertwin})
are identical and one thus gets relations like :
\begin{eqnarray}
\label{Lifj}
L_i \cdot f_{i,j}\, = \, \,f_{i,j} \cdot L_j
\end{eqnarray}

Furthermore one also verifies the following relations
 between the order one operators $L_i$'s :
\begin{eqnarray}
L_3 \cdot L_0 &=& L_8 \cdot L_2, \qquad L_4 \cdot L_0 = L_6 \cdot L_1 \\ 
L_5 \cdot L_1 &=& L_7 \cdot L_2           \nonumber   \\ 
L_{12} \cdot L_3 &=& L_{14} \cdot L_4, \qquad  L_{13} \cdot L_{17} = L_{15} \cdot L_{16}\nonumber   \\
L_{18} \cdot L_5 &=& L_{14} \cdot L_6, \qquad  L_{19} \cdot L_{23} = L_{15} \cdot L_{20} \\
L_{18} \cdot L_7 &=& L_{12} \cdot L_8, \qquad  L_{19} \cdot L_{21} = L_{13} \cdot L_{22}\nonumber   \\
L_{17} \cdot L_{11} &=& L_{22} \cdot L_9, \qquad  L_{16} \cdot L_{11} = L_{20} \cdot L_{10}\nonumber   \\
L_{23} \cdot L_{10} &=& L_{21} \cdot L_9   
\end{eqnarray}
It is straightforward to see that a relation like 
$\, L_i \cdot  L_j \, = \, \,  L_k \cdot  L_m$ yields relations like :
\begin{eqnarray}
&&L_i \cdot  (L_j\, -L_m) \, = \, \,  (L_k\, -L_i) \cdot  L_m,    \nonumber \\
&&(L_i\, -L_k) \cdot  L_j \, = \, \,  L_k \cdot  (L_m\, -L_j) , 
 \nonumber \\
&&f_{i,m} = L_j\, -L_m \, = \, \, L_k\, -L_i, \quad \quad  
f_{k,j} = L_m\, -L_j \, = \, \, L_i\, -L_k \nonumber 
\end{eqnarray}
The number of intertwiners $\, f_{i,j}$ between the operators of order one
in (\ref{Lifj}),  is at first sight
$325$, but due to the above identities only $15$ are independent.


\vskip 0.5cm

\begin{thebibliography}{99}


\bibitem{wu-mc-tr-ba-76}  T.T. Wu, B.M. McCoy, C.A. Tracy and E. Barouch,
1976 Phys. Rev. {\bf B 13}, 316-374 

\bibitem{nappi-78} C. R. Nappi, 1978  Nuovo Cim. A \textbf{44} 392

\bibitem{pal-tra-81} J. Palmer, C. Tracy, 1981 Adv. Appl. Math. \textbf{2} 329

\bibitem{yamada-84} K. Yamada, 1984 Prog. Theor. Phys. \textbf{71} 1416

\bibitem{yamada-85} K. Yamada, 1985 Phys. Lett. A \textbf{112} 456-458

\bibitem{nickel-99} B. Nickel, 1999 J. Phys. A: Math. Gen. \textbf{32} 3889-3906

\bibitem{nickel-00} B. Nickel, 2000 J. Phys. A: Math. Gen. \textbf{33}  1693-1711

\bibitem{or-ni-gu-pe-01b} W.P. Orrick, B.G. Nickel, A.J. Guttmann, J.H.H.
Perk, 2001 J. Stat. Phys. \textbf{102}  795-841

\bibitem{or-ni-gu-pe-01} W.P. Orrick, B.G. Nickel, A.J. Guttmann, J.H.H.
Perk, 2001 Phys. Rev. Lett. \textbf{86} 4120-4123


\bibitem{coy-wu-80} B.M. McCoy, T.T. Wu, 1980 Phys. Rev. Lett. \textbf{45} 675-678

\bibitem{perk-80} J.H.H. Perk, 1980 Phys. Lett. A\textbf{79} 3-5

\bibitem{jim-miw-80} M. Jimbo, T. Miwa, 1980 Proc. Japan Acad. A \textbf{56}
 405; 1981 Proc. Japan Acad. A \textbf{57}  347

\bibitem{coy-wu-81} B.M. McCoy and T.T. Wu, Nucl. Phys. \textbf{B180} (1981) 89

\bibitem{HaMa88} D. Hansel and J-M. Maillard, 1988 J. Phys. A: Math. Gen. \textbf{21} 213-225

\bibitem{ha-ma-oi-ve-87} D. Hansel, J.M. Maillard, J. Oitmaa, M.J. Vergakis,
J. Stat. Phys. 1987 \textbf{48}  69-80

\bibitem{gut-ent-96} A.J. Guttman, I.G. Enting, 1996 Phys. Rev. Lett. \textbf{76} 344-347


\bibitem{ze-bo-ha-ma-04} N. Zenine, S. Boukraa, S. Hassani, J.M. Maillard,
2004  J. Phys. A: Math. Gen. {\bf 37} 9651-9668 and  arXiv:math-ph/0407060

\bibitem{ze-bo-ha-ma-05} N. Zenine, S. Boukraa, S. Hassani, J.M. Maillard,
{\em Square lattice Ising model susceptibility: Series expansion method
and differential equation for $\chi^{(3)}$}, 2005 to be published in
J. Phys. A: Math. Gen.
 and arXiv:hep-ph/0411051

\bibitem{Meijer} J. H. Davenport,
 {\em The Difficultie od Definite Integration}, \\
http://www-calfor.lip6.fr/~rr/Calculemus03/davenport.pdf




\bibitem{Put} M. van der Put and M. F. Singer, {\em Galois theory of difference equations}, \\
 Springer-Verlag, Berlin, (1997), Lecture Notes in Mathematics. \\
$ http://www4.ncsu.edu/~singer/papers/dbook.ps$

\bibitem{Dmodule} P. H. Berman and  M. F. Singer, {\em Calculating the 
Galois Group of $\, L_1(L_2(y)) = 0$,
$\, L_1$, $\, L_2$ Completly Reducible Operators. 
},  (1998)
$http://www4.ncsu.edu/~singer/papers/12Inhom.ps$




\bibitem{Hoeij} M. Van Hoeij, {\em Formal solutions and factorization of differential 
operators with power series coefficients.}, Journal of Symbolic Computation, 
vol. 24, no1, July (1997), 1-30.


\bibitem{Weil} J-A. Weil, {\em Absolute Factorization of Differential Operators},\\
http://pauillac.inria.fr/algo/seminars/sem96-97/weil.html


\bibitem{Telescoping} S.A. Abramov, J.J. Carette, K.O. Geddes, and H.Q. Le,
  {\em Telescoping in the context of symbolic summation in Maple.}
 Journal of Symbolic Computation, 38 (4),  (2004)  1303-1326.

\bibitem{Ore} M. Giesbrecht, Y. Zhang, {\em Factoring and Decomposing Ore Polynomials over Fq(t)}.
 ACM International Symposium on Symbolic and Algebraic Computation (ISSAC), (2003) 127-134.

\bibitem{Help} M. van Hoeij,  {\em help page for   diffop},  
$\, http://www.math.fsu.edu/~hoeij/daisy/help_pages$

\bibitem{DEtools} DEtools library for differential equations \\
http://www.math.upenn.edu/ugrad/calc/manual/deplot.pdf


\bibitem{ince-56} H.K. Ince, Ordinary differential equations, (Longmans,
London, 1927; also Dover Pubs., NY, 1956)

\bibitem{Forsyth} A. R. Forsyth, Theory of Differential Equations, 6 vols.(New York: Dover,  1959)

\bibitem{gfun} Mgfun's project : see http://algo.inria.fr/chyzak/mgfun.html ; 
gfun - generating functions package see gfun in : http://algo.inria.fr/libraries 



\bibitem{Singer} Michael F. Singer,  {\em Testing reducibility of linear
 differential operators: A group theoretic perspective.},  Applicable Algebra 
in Engineering, Communication and Computing, vol. 7, no2, (1996), 77-104.





\end{thebibliography}
\end{document}